\documentclass[graybox]{svmult}
\usepackage{mathptmx}       % selects Times Roman as basic font
\usepackage{helvet}         % selects Helvetica as sans-serif font
\usepackage{courier}        % selects Courier as typewriter font
\usepackage{type1cm}        % activate if the above 3 fonts are
\usepackage{makeidx}         % allows index generation
\usepackage{graphicx}        % standard LaTeX graphics tool
\usepackage{multicol}        % used for the two-column index
\usepackage[bottom]{footmisc}% places footnotes at page bottom
\makeindex             % used for the subject index
\begin{document}

\title*{Observed Consequences of Presupernova Instability in Very
  Massive Stars} 
\titlerunning{Observed Deaths of Very Massive Stars} 
\author{Nathan Smith}
% Use \authorrunning{Short Title} for an abbreviated version of
% your contribution title if the original one is too long
\institute{Nathan Smith \at Steward Observatory, 933 N.\ Cherry Ave.,
  Tucson, AZ 85721, USA, \email{nathans@as.arizona.edu}}
% \and Name of Second Author \at Name, Address of Institute
% \email{name@email.address}}
%
% Use the package "url.sty" to avoid
% problems with special characters
% used in your e-mail or web address
\maketitle

%1)  Empirical properties of VMS (Martins confirmed)
%2)  Massive star Formation Theory (Krumholz confirmed)
%3)  Mass Loss (Vink)
%4)  Envelope instabilities (Owocki)
%5)  Evolution + Fate (Heger & Woosley confirmed)
%6)  explosion properties (Nathan Smith)

%\abstract*{Each chapter should be preceded by an abstract (10--15
%  lines long) that summarizes the content. The abstract will appear
%  \textit{online} at \url{www.SpringerLink.com} and be available with
%  unrestricted access. This allows unregistered users to read the
%  abstract as a teaser for the complete chapter. As a general rule the
%  abstracts will not appear in the printed version of your book unless
%  it is the style of your particular book or that of the series to
%  which your book belongs.  Please use the 'starred' version of the
%  new Springer \texttt{abstract} command for typesetting the text of
%  the online abstracts (cf. source file of this chapter template
%  \texttt{abstract}) and include them with the source files of your
%  manuscript. Use the plain \texttt{abstract} command if the abstract
%  is also to appear in the printed version of the book.}

\abstract{This chapter concentrates on the deaths of very massive
  stars, the events leading up to their deaths, and how mass loss
  affects the resulting death.  The previous three chapters emphasized
  the theory of wind mass loss, eruptions, and core collapse physics,
  but here we emphasize mainly the observational properties of the
  resulting death throes.  Mass loss through winds, eruptions, and
  interacting binaries largely determines the wide variety of
  different types of supernovae that are observed, as well as the
  circumstellar environments into which the supernova blast waves
  expand.  Connecting these observed properties of the explosions to
  the initial masses of their progenitor stars is, however, an
  enduring challenge and is especially difficult for very massive
  stars.  Superluminous supernovae, pair instability supernovae, gamma
  ray bursts, and ``failed'' supernovae are all end fates that have
  been proposed for very massive stars, but the range of initial
  masses or other conditions leading to each of these (if they
  actually occur) are still very certain.  Extrapolating to infer the
  role of very massive stars in the early universe is essentially
  unencumbered by observational constraints and still quite dicey.}

\section{Introduction}
\label{sec:1}

As discussed in previous chapters (Vink, Owocki), two critical aspects
in the evolution of very massive stars (VMSs) are that their high
luminosities cause strong mass loss in radiation-driven winds, and
that high luminosities can also cause severe instabilities in the
stellar envelope and interior as the star approaches the Eddington
limit.  These features become increasingly important as the initial
stellar mass increases, but especially so as the star evolves off the
main sequence and approaches its death. Moreover, the two are
interconnected, since mass loss will increase the star's
luminosity/mass ratio, possibly leading to more intense instabilities
over time.

It should not be surprising, then, that VMSs show clear empirical
evidence of this instability, and this chapter discusses various
observational clues that we have.  This is a particularly relevant
topic, as time-domain astronomy is becoming an increasingly active
field of observational research.  Throughout, the reader should
remember that we are focussed on observed phenomena, and that working
backward to diagnose possible underlying physical causes is not always
straightforward.  Hence, this interpretation is where most of the
current speculation and debate rests among researchers working in the
field.  Stellar evolution models make predictions for the appearance
of single massive stars late in their lives, but the influence of
binary interaction may be extremely important or even dominant (Langer
2012), and the assumptions about mass-loss that go into the
single-star models are not very reliable (Smith 2014).  In particular,
the eruptive instabilities discussed in this chapter are not included
in single-star evolution models, and as such, these models provide us
with little perspective for understanding the very latest unstable
phases of VMSs or their final fates.  The loosely bound envelopes that
result from a star being close to the Eddington limit may be an
important factor in directly causing outbursts, but having a barely
bound envelope may also make it easier for other mechanisms to be
influential, such as energy injection from non-steady nuclear burning,
precursor core explosions, or binary interactions (see e.g., Smith \&
Arnett 2014 for a broader discussion of this point).

In the sections to follow, we discuss the observed class of eruptive
luminous blue variables (LBVs) that have been linked to the late
evolutionary phases of VMSs, various types of very luminous supernovae
(SNe) or other explosions that may come from VMSs, and direct
detections of luminous progenitors of SNe (including a few actual
detections of pre-SN eruptions) that provide a link between VMSs and
their SNe.

\section{LBVs and their Giant Eruptions} %- 10 pages
\label{sec:2}

Perhaps the most recognizable manifestation of the instability that
arises in the post-main-sequence evolution of VMSs is the class of
objects known as luminous blue variables (LBVs).  These were
recognized early-on as the brightest blue irregular variables in
nearby galaxies like M31, M33, and NGC~2403 (Hubble \& Sandage 1953;
Tammann \& Sandage 1968), and these classic examples were referred to
as the ``Hubble-Sandage variables''.  Later, Conti (1984) recognized
that many different classes of hot, irregular variable stars in the
Milky Way and Magellanic clouds were probably related to these
Hubble-Sandage variables, and probably occupy similar evolutionary
stages in the lives of massive stars, so he suggested that they be
grouped together and coined the term ``LBV'' to describe them
collectively. The LBVs actually form a rather diverse class,
consisting of a wide range of irregular variable phenomena associated
with evolved massive stars (see reviews by Humphreys \& Davidson 1994;
van Genderen 2001; Smith et al.\ 2004, 2011a; Van Dyk \& Matheson
2012; Clark et al.\ 2005).

\subsection{Basic observed properties of LBVs}

In addition to their high luminosities, some of the key observed
characteristics of LBVs are as follows (although beware that not all
LBVs exhibit all these properties):

\begin{itemize}

\item {\bf S Doradus eruptions.} Named after the prototype in the LMC,
  S Dor eruptions are seen as a brightening that occurs at visual
  wavelengths resulting from a change in apparent temperature of the
  star's photosphere; this causes the peak of the energy distribution
  to shift from the UV to visual wavelengths at approximately constant
  bolometric luminosity.  The increase in visual brightness (i.e. 1--2
  mag, typically for more luminous stars) corresponds roughly to the
  bolometric correction for the star, so that hotter stars exhibit
  larger amplitudes in their S Dor events.  LBVs have different
  temperatures in their quiescent state, and this quiescent
  temperature increases with increasing luminosity.  The visual
  maximum of S Dor eruptions, on the other hand, usually occurs at a
  temperature around 7500 K regrdless of luminosity, causing the star
  to resemble a late F-type supergiant with zero bolometric correction
  (see Figure~\ref{fig:hrd}).  While these events are defined to occur
  at constant bolometric luminosity (Humphreys \& Davidson 1994), in
  fact quantitative studies of classic examples like AG Car do reveal
  some small varition in $L_{Bol}$ through the S Dor cycle (Groh et
  al.\ 2009). Similarly, the traditional explanation for the origin of
  the temperature change was that the star increases its mass-loss
  rate, driving the wind to very high optical depth and the creation
  of a pseudo photosphere (Humphreys \& Davidson 1994; Davidson 1987).
  Quantitative spectroscopy reveals, however, that the measured
  mass-loss rates do not increase enough to cause a pseudo photosphere
  in classic S Dor variables like AG Car (de Koter et al.\ 1996), and
  that the increasing photospheric radius is therefore more akin to a
  true expansion of the star's photosphere (i.e., a pulsation).
  Possible causes of this inflation of the star's outer layers is
  discussed elsewhere in this book (see Owocki's chapter).  LBVs that
  experience these excursions are generally thought to be very massive
  stars, but their mass range is known to extend down to around 25
  $M_{\odot}$ (Smith et al.\ 2004).

\begin{figure}[b]
%\sidecaption
%% Use the relevant command for your figure-insertion program
%% to insert the figure file.
%% For example, with the graphicx style use
%\includegraphics[scale=0.67]{hrdfig1.eps}
\includegraphics[scale=0.62]{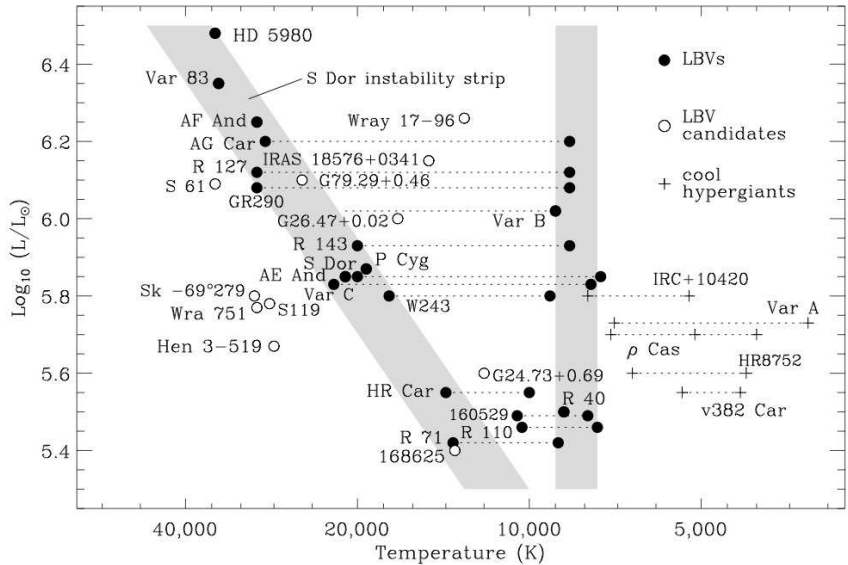}
%%
%% If no graphics program available, insert a blank space i.e. use
%%\picplace{5cm}{2cm} % Give the correct figure height and width in cm
%%
\caption{The upper HR Diagram of LBVs and some LBV candidates (from
  Smith, Vink, \& de Koter 2004).  The most massive LBVs and LBV
  candidates like $\eta$ Car and the Pistol star are off the top of
  this diagram.  The diagonal strip where LBVs reside at quiescence is
  the S Dor instability strip discussed in the text.  Note that LBVs
  are recognized by their characteristic photometric variability down
  to luminosities where the S Dor instability strip meets the
  eruptive temperature.}
\label{fig:hrd}       % Give a unique label
\end{figure}

\item {\bf Quiescent LBVs reside on the S Dor instability strip.}  As
  noted in the previous point, LBVs all show roughly the same apparent
  temperature in their cool/bright state during an outburst, but they
  have different apparent temperatures in their hot/quiescent states.
  These hot temperatures are not random.  In quiescence, most LBVs
  reside on the so-called ``S Dor instability strip'' in the HR
  Diagram (Wolf 1989).  This is a diagonal strip, with increasing
  temperature at higher luminosity (see Figure~\ref{fig:hrd}).
  Notable examples that do not reside on this strip are the most
  luminous LBVs, like $\eta$ Car and the Pistol star, so the S Dor
  instability strip may not continue to the most massive and most
  luminous stars, for reasons that may be related to the strong winds
  in these VMSs (see Vink chapter).  Many of the stars at the more
  luminous end of the S Dor instability strip are categorized as
  Ofpe/WN9 or WNH stars in their hot/quiescent phases, with AG Car and
  R127 being the classic examples where these stars are then observed
  to change their spectral type and suffer bona-fide LBV outbursts.
  There are also many Ofpe/WN9 stars in the same part of the HR
  Diagram that have not exhibited the characteristic photometric
  variability of LBVs in their recent history, but which have
  circumstellar shells that may point to previous episodes of eruptive
  mass loss (see below).  Such objects with spectroscopic similarity
  to quiescent LBVs, but without detection of their photometric
  variability, are sometimes called ``LBV candidates''.

\item {\bf Giant eruptions.}  The most dramatic variability attributed
  to LBVs is the so-called ``giant eruptions'', in which stars are
  observed to increase their radiative luminosity for months to years,
  accompanied by severe mass loss (e.g., Humphreys et al.\ 1999; Smith
  et al.\ 2011a). The star survives the disruptive event. The best
  studied example is the Galactic object $\eta$ Carinae, providing us
  with its historically observed light curve (Smith \& Frew 2011), as
  well as its complex ejecta that contain 10-20 $M_{\odot}$ and
  $\sim$10$^{50}$ ergs of kinetic energy (Smith et al. 2003; Smith
  2006).  Besides the less well-documented case of P Cygni's 1600 AD
  eruption, our only other examples of LBV-like giant eruptions are in
  other nearby galaxies. A number of these have been identified, with
  peak luminosities similar to $\eta$ Car or less (Van Dyk \& Matheson
  2012; Smith et al. 2011a). Typical expansion speeds in the ejecta
  are 100 - 1000 km s$^{-1}$ (Smith et al. 2011a). These events are
  discussed more below.

\item {\bf Strong emission-line spectra.} Most, but not all, LBVs
  exhibit strong emission lines (especially Balmer lines) in their
  visual-wavelength spectra.  This is a consequence of their very
  strong and dense stellar winds (see Vink chapter), combined with
  their high UV luminosity and moderately high temperature.  The wind
  mass-loss rates implied by quantitative models of the spectra range
  from 10$^{-5}$ to 10$^{-3}$ $M_{\odot}$ yr$^{-1}$; this is enough to
  play an important role in the evolution of the star (see Smith
  2014), and eruptions enhance the mass loss even more. The emission
  lines in LBVs are, typically, much stronger than the emission lines
  seen in main-sequence O-type stars of comparable luminosity, and all
  of the more luminous LBVs have strong emission lines.  Other stars
  that exhibit similar spectra but are not necessarily LBVs include
  WNH stars, Ofpe/WN9 stars, and B[e] supergiants, some of which
  occupy similar parts of the HR Diagram.

\item {\bf Circumstellar shells.}  Many LBVs are surrounded by
  spatially resolved circumstellar shells.  These fossil shells
  provide evidence of a previous eruption.  Consequently, some stars
  that resemble LBVs spectroscopically and have massive circumstellar
  shells, but have not (yet) been observed to exhibit photometric
  variability characteristic of LBVs, are often called LBV candidates.
  Many authors prefer to group LBVs and LBV candidates together (the
  logic being that a volcano is still a volcano even when it is
  dormant). LBV circumstellar shells are extremely important, as they
  provide the only reliable way to estimate the amount of mass ejected
  in an LBV giant eruption.  The most common technique for measuring
  the mass is by calculating a dust mass from thermal-IR radiation,
  and then converting this to a total gas mass with an assumed
  gas:dust mass ratio (usually taken as 100:1, although this value is
  uncertain\footnote{If this value is wrong, it is probably a
    conservative underestimate.  This is because a gas:dust mass ratio
    of 100:1 assumes that all refractory elements at $Z_{\odot}$ are
    in grains, whereas in reality, the dust formation may be less
    efficient or UV and shocks may destroy some dust, leaving some of
    these elements in the gas phase (and thus raising the total mass).
    In general, nebular gas masses inferred from thermal-IR dust
    emission should be considered lower limits, especially at $Z <
    Z_{\odot}$.}).  To calculate a dust mass from the IR luminosity,
  one must estimate the dust temperature from the spectral energy
  distribution (SED), and then adopt some wavelength-dependent grain
  opacities in order to calculate the emitting mass.  The technique
  can be quite sensitive to multiple temperature components, and
  far-IR data have been shown to be very important because most of the
  mass can be hidden in the coolest dust, which is often not
  detectable at wavelengths shorter than 20 $\mu$m.  One can also
  measure the gas mass directly by various methods, usually adopting a
  density diagnostic like line ratios of [Fe~{\sc ii}] or [S~{\sc ii}]
  and multiplying by the volume and filling factor, or calculating a
  model for the density needed to produce the observed ionization
  structure using codes such as CLOUDY (Ferland et al.\ 1998). The
  major source of uncertainty here is the assumed ionization
  fraction. Masses of LBV nebulae occupy a very large range from
  $\sim$20 $M_{\odot}$ at the upper end down to 0.1 $M_{\odot}$ (Smith
  \& Owocki 2006), although even smaller masses become difficult to
  detect around bright central stars.

\item{\bf Wind speeds and nebular expansion speeds.}  LBV winds and
  nebulae typically have expansion speeds of 50-500 km s$^{-1}$, due
  to the fact that the escape speed of the evolved blue supergiant is
  lower than for the more compact radii of O-type stars and WR stars
  that have faster speeds of order 1000 km s$^{-1}$.  In many cases,
  the shell nebulae are expanding with an even slower speed than the
  underlying wind, but this is not always the case.  The slower
  nebular speeds may suggest that the nebulae were ejected in a state
  when the star was close to the Eddington limit (lower effective
  gravity) or that the LBV eruption ejecta have decelerated after
  colliding with slow CSM or high-pressure ISM.

\item {\bf N-rich ejecta.} Lastly, LBVs typically exhibit strong
  enhancements in their N abundance, measured in the circumstellar
  nebulae or in the wind spectrum.  The most common measurement
  involves the analysis of visual-wavelength spectra, using nebular
  [S~{\sc ii}] lines to derive an electron density, using the [N~{\sc
    ii}] ($\lambda$6583+$\lambda$6548)/$\lambda$5755 ratio to derive
  an electron temperature, and then using the observed intensity of
  the [N~{\sc ii}] lines compared to H lines for a relative N$^+$/H
  ratio, and then doing a similar analysis of O and C lines in order
  to estimate N/O and N/C ratios.  One must make assumptions about the
  ionization levels of N and other elements, but if UV spectra are
  available, one can constrain the strength of a wide range of
  ionization levels of each atom.  In the case of $\eta$ Carinae, for
  example, strong lines of N~{\sc i}, {\sc ii}, {\sc iii}, {\sc iv},
  and {\sc v} are detected, but O lines of all ionization levels are
  extremely faint (Davidson et al.\ 1986).  The observed levels of N
  enrichment in LBVs suggest that the outer layers of the stars
  include large quantities of material processed through the CNO cycle
  and mixed to the surface, requiring that LBVs are post-main-sequence
  stars.

\end{itemize}

\subsection{The Evolutionary State of LBVs}

While evidence for N enrichment and C+O depletion suggest that LBVs
are massive post-main-sequence stars, their exact evolutionary status
within that complex and possibly non-monotonic evolution has been
controversial - moreso in recent years.

The traditional view of LBVs, which emerged in the 1980s and 1990s, is
that they correspond to a very brief transitional phase of massive
star evolution, as the star moves from core H burning when it is seen
as a main sequence O-type star, to core He burning when it is seen as
a Wolf-Rayet (WR) star.  A typical monotonic evolutionary scheme for a
VMS is as follows:

\smallskip 

\noindent 100 $M_{\odot}$: O star $\rightarrow$ Of/WNH $\rightarrow$
LBV $\rightarrow$ WN $\rightarrow$ WC $\rightarrow$ SN Ibc

\smallskip

\noindent In this scenario, the strong mass-loss experienced by LBVs
is important for removing what is left of the star's H envelope after
the main sequence, leaving a hydrogen-poor WR star following the end
of the LBV phase.  The motivation for thinking that this is a very
brief phase comes from the fact that LBVs are extremely rare, even for
very massive stars: taking the relative numbers of LBVs and O-type
stars at high luminosity, combined with the expected H-burning
lifetime of massive O-type stars, would imply a duration for the LBV
phase of only a few 10$^4$ yr or less.  This view fits in nicely with
a scenario where the observed population of massive stars is dominated
by single-star evolution.

However, a number of problems and inconsistencies have arisen with
this standard view of LBVs.  For one thing, the very short
transitional lifetime depends on the assumption that the observed LBVs
are representative of the whole transitional phase.  In fact, there is
a much larger number of blue supergiant stars that are not {\it bona
  fide} LBVs seen in eruption, but which are probably related ---
these are the LBV candidates discussed earlier.  Examining populations
in nearby galaxies, for example, Massey et al. (2007) find that there
are more than {\it an order of magnitude} more LBV candidates than
there are LBVs confirmed by their photometric variability.  (For
example, there are several hundred LBV candidates in M31 and M33,
compared to the 8 LBVs known by their photometric variability.)  If
the LBV candidates are included with LBVs, then the average lifetime
of the LBV phase must rise from a few 10$^4$ yr to several 10$^5$ yr.
Now we have a problem, because this is a significant fraction of the
whole He burning lifetime, making it impossible for LBVs to be
fleeting {\it transitional} objects.  There is not enough time in
core-He burning to link them to both WR stars and LBVs. Should we
include the LBV candidates and related stars? Are they dormant LBVs?
If indeed LBVs go through dormant phases when they are not showing
their instability (or when they have temporarily recovered from the
instability after strong mass loss), then it would be a mistake not to
include the duty cycle of instability in the statistics of LBVs.
Massey (2006) has pointed to the case of P Cygni as a salient example:
its 1600 A.D.\ giant LBV eruption was observed and so we consider it
an architypal LBV, but it has shown no eruptive LBV-like behavior
since then. If the observational record had started in 1700, then we
would have no idea that P Cygni was an LBV and we would be wrong.  So
how many of the other LBV candidates are dormant LBVs?  The massive
circumstellar shells seen around many LBV candidates imply that they
have suffered LBV giant eruptions in the previous 10$^3$ yr or so.

Another major issue is that we have growing evidence that LBVs or
something like them (massive stars with high mass loss, N enrichment,
H rich, slow $\sim$100 km s$^{-1}$ winds, massive shells) are
exploding as core-collapse SNe while still in an LBV-like phase (see
below).  This could not be true if LBVs are only in a brief transition
to the WR phase, which should last another 0.5-1 Myr before core
collapse to yield a SN Ibc.  Pre-supernova eruptive stars that
resemble LBVs are discussed in more detail in following sections.

Last, the estimates for lifetimes in various evolutionary phases in
the typical monotonic single-star scenario (see above) ignore
empirical evidence that binary evolution dominates the evolution of a
large fraction of massive stars.  Many massive O-type stars (roughly
1/2 to 2/3) are in binary systems whose orbital separation is small
enough that they should interact and exchange mass during their
lifetime (Kobulnicky \& Fryer 2007; Kiminki \& Kobulnicky 2012;
Kiminki et al.\ 2012; Chini et al.\ 2012; Sana et al.\ 2012).  These
binary systems {\it must} make a substantial contribution to the
observed populations of evolved massive stars and SNe, so to find
agreement between predictions of single-star evolutionary models and
observed populations indicates that something is wrong with the models.
Unfortunately, solutions to these problems are not yet readily
apparent; some current effort is focussed here, and these topics are
still a matter of debate among massive star researchers.

\subsection{A special case: Eta Carinae}

The enigmatic massive star $\eta$ Carinae is perhaps the most famous
and recognizable example of an evolved and unstable VMS.  It is
sometimes regarded as the prototype of eruptive LBVs, but at the same
time it has a long list of peculiarites that make it seem unique and
very atypical of LBVs.  In any case, it is by far the {\it best
  studied} LBV, and (for better or worse) it has served as a benchmark
for understanding LBVs and the physics of their eruptions.

Several circumstances conspire to make $\eta$ Car such a fountain of
information.  It is nearby (about 2.3 kpc; Smith 2006) and bright with
low interstellar extinction, so one is rarely photon-starved when
observing this object at any wavelength.  It is one of the most
luminous and massive stars known, with rough values of $L \simeq 5
\times 10^6$ $L_{\odot}$ and a present-day mass for the primary around
100 $M_{\odot}$ (its ZAMS mass is uncertain, but was probably a lot
more than this).  Its giant eruption in the 19th century was observed
at visual wavelengths so that we have a detailed light curve of the
event (Smith \& Frew 2011), and $\eta$ Car is now surrounded by the
spectacular expanding Homunculus nebula that provides us with a fossil
record of that mass loss.  This nebula allows us to estimate the mass
and kinetic energy of the event, which are $\sim$15 $M_{\odot}$ and
$\sim$10$^{50}$ erg, and we can measure the geometry of the mass
ejection because the Homunculus is still young and in free expansion
(Smith et al.\ 2003; Smith 2006).

\begin{figure}[b]
\includegraphics[scale=0.59]{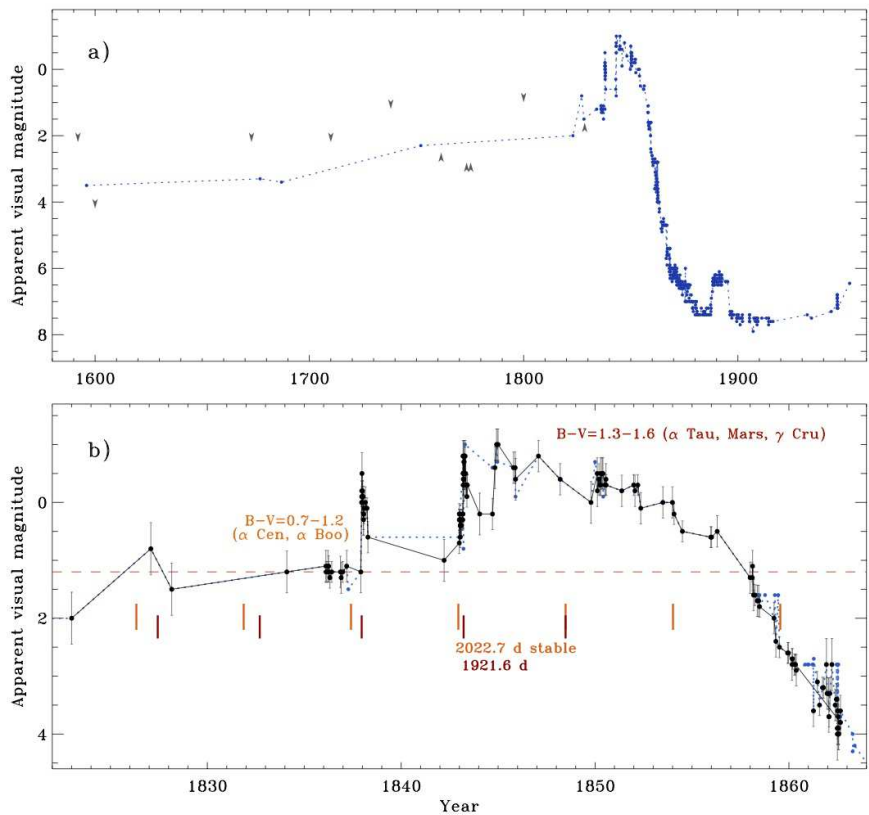}
\caption{The historical light curve of the 19th century Great Eruption
  of $\eta$ Carinae, from Smith \& Frew (2011).}
\label{fig:etaLC} 
\end{figure}

Davidson \& Humphreys (1997) provided a comprehensive review of the
star and its nearby ejecta in the mid-1990s, but there have been many
important advances in the subsequent 16 years.  It has since been well
established that $\eta$~Car is actually in a binary system with a
period of 5.5 yr and $e \simeq 0.9$ (Damineli et al.\ 1997), which
drastically alters most of our ideas about this object.  Accordingly,
much of the research in the past decade has been devoted to
understanding the temporal variability in this colliding-wind binary
system (see Madura et al.\ 2012, and references therein).  Detailed
studies of the Homunculus have constrained its 3D geometry and
expansion speed to high precision (Smith 2006), and IR wavelengths
established that the nebula contains almost an order of magnitude more
mass than was previously thought (Smith et al.\ 2003; Morris et al.\
1999; Gomez et al.\ 2010).  The larger mass and kinetic energy force a
fundamental shift in our undertanding of the physics of the Great
Eruption (see below).  Observations with {\it HST} have dissected the
detailed ionization structure of the nebula and measured its expansion
proper motion (e.g., Gull et al.\ 2005, Morse et al.\ 2001).  Spectra
have revealed that the Great Eruption also propelled extremely fast
ejecta and a blast wave outside the Homunculus, moving at speeds of
5000 km s$^{-1}$ or more (Smith 2008).  We have an improved record of
the 19th century light curve from additional released historical
documents (Smith \& Frew 2011; Figure~\ref{fig:etaLC}), and perhaps
most exciting, we have now detected light echoes from the 19th century
eruption, allowing us to obtain spectra of the outburst itself after a
delay of 160 years (Rest et al.\ 2012).

Altogether, the outstanding observational record of $\eta$ Car
suggests a picture wherein a VMS suffered an extremely violent,
$\sim$10$^{50}$ erg explosive event comparable to a weak supernova,
which ejected much of the star's envelope - but the star apparently
survived this event.  This gives us a solid example of the extreme
events that can result from the instability in a VMS, but the
underlying physics is still not certain.  Interactions with a close
companion star are critical for understanding its present-day
variability; the binary probably played a critical role in the
behavior of the 19th century Great Eruption as well, although the
details are still unclear.

While $\eta$ Car is the best observed LBV, it may not be very
representative of the LBV phenomenon in general.  In what ways is
$\eta$ Car so unusual among LBVs?  Its 19th century Great Eruption
reached a similar peak absolute magnitude ($-$14 mag) to those of
other so-called ``SN impostor'' events in nearby galaxies (see Smith
et al.\ 2011a), but unlike most extragalactic examples, its eruption
persisted for a decade or more, whereas most extragalactic examples of
similar luminosity last only 100 days or less.  Among well-studied
LBVs in the Galaxy and Magellanic Clouds, only $\eta$ Car is known to
be in a wide colliding-wind binary system that shows very pronounced,
slow periodic modulation across many wavelengths (HD~5980 in the SMC
is in a binary, but with a much shorter period).  Its 500 km s$^{-1}$
and 10$^{-3}$ $M_{\odot}$ yr$^{-1}$ wind is unusually fast and dense
compared to most LBVs, which are generally an order of magnitude less
dense.  Its Homunculus nebula is the youngest LBV nebula, and together
with P Cygni these are the only sources for which we have both an
observed eruption event and the nebula it created.  Thus, it remains
unclear if $\eta$ Car represents a very brief (and therefore rarely
observed) violent eruption phase that most VMSs pass through at some
time in their evolution, or if it really is so unsual because of its
very high mass and binary system parameters.

In any case, the physical parameters of $\eta$ Car's eruption are
truly extreme, and they push physical models to limits that are
sometimes hard to meet.  The 19th century event has long been the
protoype for a super-Eddington wind event, but detailed investigation
of the physics involved shows that this is quite difficult to achieve
(see Owocki's chapter in this volume).  At the same time, we now have
mounting observational evidence of an explosive nature to the Great
Eruption: (1) A very high ratio of kinetic energy to integrated
radiated energy, exceeding unity; (2) Brief spikes in the light curve
that occur at times of periastron; (3) evidence for a small mass of
very fast moving ($\sim$5000 km s$^{-1}$) ejecta and a blast wave
outside the Homunculus, which requires a shock-powered component to
the eruption, and (4) behavior of the spectra seen in light echoes,
which do not evolve as expected from an opaque wind.  These hints
suggest that some of the phenomena we associate with LBVs (and their
extragalactic analogs) are driven by explosive physics
(i.e. hydrodynamic events in the envelope) rather than (or in addition
to) winds driven from the surface by high luminosity.  This is
discussed in more detail in the following subsection.

\subsection{Giant Eruptions: Diversity, Explosions, and Winds}

Giant eruptions are simultaneously the most poorly understood, most
puzzling, and probably the physically most important of the observed
phenomena associated with LBVs.  They are potentially the most
important aspect for massive stars because of the very large amounts
of mass (as much as 10-20 $M_{\odot}$) that are ejected in a short
amount of time, and consequently, because of their dramatic influence
on immediate pre-SN evolution (next section).  Although the giant
eruptions themselves are rarely observed because they are infrequent
and considerably fainter than SNe, a large number of LBVs and
spectroscopically similar stars in the Milky Way and Magellanic Clouds
are surrounded by massive shell nebulae, indicating previous eruptions
with a range of ejecta masses from 1-20 $M_{\odot}$ (Clark et al.\
2005; Smith \& Owocki 2006; Wachter et al.\ 2010; Gvaramadze et al.\
2010). Thus, eruptive LBV mass loss is inferred to be an important
effect in late evolution of massive stars, and perhaps especially so
in VMSs.

Originally the class of LBV giant eruptions was quite exclusive, with
only four approved members: $\eta$ Car's 1840s eruption, P Cygni's
1600 AD eruption, SN~1954J (V12 in NGC~2403), and SN~1961V (see
Humphreys, Davidson, \& Smith 1999).  Due the the advent of dedicated
searches for extragalactic SNe from the late 1990s onward, the class
of giant eruptions has grown to include a few dozen members (see
recent summaries by Smith et al.\ 2011a; Van Dyk \& Matheson 2012).
Because of their serendipitous discovery in SN searches, they are also
referred to as ``SN impostors''.  Other names include ``Type V''
supernovae (from F.\ Zwicky), ``$\eta$ Car analogs'', and various
permutations of ``intermediate luminosity transients''.

Although the total number of SN impostors is still quite small
(dozens) compared to SNe (thousands), the actual rates of these events
could potentially be comparable to or even exceed those of
core-collapse SNe.  The difference is due to the fact that by
definition, SN impostors are considerably fainter than true SNe, and
are therefore much harder to detect.  Since they are $\sim$100 times
less luminous than a typical Type Ia SN, their potential discovery
space is limited to only 1/1000 of the volume in which SNe can be
discovered with the same telescope.  Their discovery is made even more
difficult because of the fact that their contrast compared to the
underlying host galaxy light is lower, and because in some cases they
have considerably longer timescales and much smaller amplitudes of
variability than SNe.  Unfortunately, there has not yet been any
detailed study of the rates of SN impostors corrected for the inherent
detection bias in SN searches.  We are limited to small numbers, but
one can infer that the rates of LBV eruptions and core-collapse SNe
are comparable based on a local guesstimate: in our nearby region of
the Milky Way there have been 2 giant LBV eruptions (P Cyg \& $\eta$
Car) and 3 SNe (Tycho, Kepler, and Cas A; and only 1 of these was a
core-collapse SN) in the past $\sim$400 yr.

The increased number of SN impostors in the past decade has led to
recognition of wide diversity among the group, and correspondingly,
increased ambiguity about their true physical nature.  It is quite
possible that many objects that have been called ``SN impostors'' are
not LBVs, but something else.  The SN impostors have peak absolute
magnitudes around $-$14 mag, but there is actually a fairly wide
spread in peak luminosity, ranging from $-$15 mag down to around $-$10
mag. At higher luminosity, transients are assumed to be supernovae,
and at lower luminosity we call them something else (novae, stellar
mergers, S Dor eruptions, etc.) --- but these dividing lines are
somewhat arbitrary.  Most of their spectra are similar, the most
salient characteristic being bright, narrow H emission lines (so they
are all ``Type IIn'') atop either a smooth blue continuum or a cooler
absorption-line spectrum.  Since the outbursts all look very similar,
many different types of objects might be getting grouped together by
observers.  When more detailed pre-eruption information about the
progenitor stars is available, however, we find a range of cases.
Some are indeed very luminous, blue, variable stars; but some are not
so luminous ($<$10$^5$ $L_{\odot}$), and are sometimes found among
somewhat older stellar populations than one expects for a VMS (Prieto
et al.\ 2008a, 2008b; Thompson et al.\ 2009).  Some well-studied
extragalactic SN impostors that are clearly massive stars suffering
LBV-like giant eruptions are SN~1997bs, SN~2009ip, UGC~2773-OT,
SN~1954J, V1 in NGC~2366, SN~2000ch; some well-studied objects that
appear to be lower-mass stars (around 6-10 $M_{\odot}$) are SN~2008S,
NGC~300-OT, V838 Mon, and SN~2010U.  There are many cases in between
where the interpretation of observational data is less straightforward
or where the data are less complete.  In any case, it is interesting
that even lower mass stars (8--15 $M_{\odot}$) may be suffering
violent eruptive instabilities similar to those seen in the most
massive stars.  If the physical cause of the outbursts is at all
related, it may point to a deep-seated core instability associated
with nuclear burning or some binary collision/merger scenario, rather
than an envelope instability associated with the quiescent star being
near the Eddington limit.

Physically, the difference between a ``SN impostor''/giant eruption
and a true (but underluminous) SN is that the star does not survive
the latter type of event.  Observationally, it is not always so easy
to distinguish between the two.  Even if the star survives, it may
form dust that obscures the star at visual wavelengths, while IR
observations may not be available to detect it.  On the other hand,
even if the star dies, there may appear to be a ``surviving'' source
at the correct position if it is a host cluster, a companion star, an
unrelated star superposed at the same position, or ongoing CSM
interaction from the young SN remnant.  It is often difficult to find
decisive evidence in the faint, noisy, unresolved smudges one is
forced to interpret when dealing with extragalactic examples.
Consider the extremely well-observed case of SN~1961V.  This object
was one of the original ``Type V'' SNe and a prototype of the class of
LBV giant eruptions (Humphreys et al.\ 1999).  However, two recent
studies have concluded that it was most likely a true core-collapse
Type IIn SN, and for two different reasons: Smith et al.\ (2011a)
point out that all of the observed properties of the rather luminous
outburst are fully consistent with the class of Type~IIn SNe, which
did not exist in 1961 and was not understood until recently.  If
SN~1961V were discovered today, we would undoubtedly call it a true
SN~IIn since its high peak luminosity ($-$18 mag) and other observed
properties clearly make it an outlier among the SN impostors. On the
other hand, Kochanek et al.\ (2011) analyzed IR images of the site of
SN~1961V and did not find an IR source consistent with a surviving
luminous star that is enshrouded by dust, like $\eta$ Car.  Both
studies conclude that since the source is now at $\sim$6 mag fainter
than the luminous blue progenitor star, it probably exploded as a
core-collapse event.  Although there is an H$\alpha$ emission line
source at the correct position (Van Dyk et al.\ 2002), this could be
due to ongoing CSM/shock interaction, since no continuum emission is
detected.  It is hard to prove definitively that the star is dead,
however (for an alternative view, see Van Dyk \& Matheson 2012).  This
question is very important, though, because the progenitor of SN~1961V
was undoubtedly a very luminous star with a likely initial mass well
exceeding 100 $M_{\odot}$.  If it was a true core-collapse SN, it
would prove that some very massive stars do explode and make
successful SNe.

What is the driving mechanism of LBV giant eruptions?  What is their
source of luminosity and kinetic energy? Even questions as simple and
fundamental as these have yet to find answers.  Two broad classes of
models have developed: super-Eddington winds, and explosive mass loss.
Both may operate at some level in various objects.

Traditionally, LBV giant eruptions have been discussed as
super-Eddington winds driven by a sudden and unexplained increase in
the star's bolometric luminosity (Humphreys \& Davidson 1994; Shaviv
2000; Owocki et al.\ 2004; Smith \& Owocki 2006), but there is growing
evidence that some of them are non-terminal hydrodynamic ejections
(see Smith 2008, 2013).  Part of the motivation for this is based on
detailed study of $\eta$ Carinae, which as noted above, has shown
several signs that the 1840s eruption had a shock-driven component to
it.  One normally expects sudden, hydrodynamic events to be brief
(i.e., a dynamical time), which may seem incongruous with the 10 yr
long Great Eruption of $\eta$ Car.  However, as in some very
long-lasting core-collapse SNe, it is possible to power the observed
luminosity of the decade-long Great Eruption with a shock wave plowing
through dense circumstellar gas (Smith 2013). In this model, the
duration of the transient brightening event is determined by how long
it takes for the shock to overrun the CSM (this, in turn, depends on
the relative speeds of the shock and CSM, and the radial extent of the
CSM). Since shock/CSM interaction is such an efficient way to convert
explosion kinetic energy into radiated luminosity, it is likely that
many of the SN impostors with narrow emission lines are in fact
powered by this method.  The catch is that even this method requires
something to create the dense CSM into which the shock expands.  This
may be where super-Eddington winds play an important role.  The
physical benefit of this model is that the demands on the
super-Eddington wind are relaxed to a point that is more easily
achievable; instead of driving 10 $M_{\odot}$ in a few years (as for
$\eta$ Car), the wind can provide roughly half the mass spread over
several decades or a century.  The required mass-loss rates are then
of order 0.01-0.1 $M_{\odot}$ yr$^{-1}$, which is more reasonable and
physically plausible than a few to several $M_{\odot}$ yr$^{-1}$.
Also, the wind can be slow (as we might expect for super-Eddington
winds; Owocki et al.\ 2004), whereas the kinetic energy in observed
fast LBV ejecta can come from the explosion.

In any case, the reason for the onset of the LBV eruption remains an
unanswered question.  In the super-Eddington wind model, even if the
wind can be driven at the rates required, we have no underlying
physical explanation for why the star's bolometric luminosity suddenly
increases by factors of 5-10 or more.  In the explosion model, the
reason for an explosive event preceding core collapse is not known,
and the cause of explosive mass loss at even earlier epochs is very
unclear.  It could either be caused by some instability in late
nuclear burning stages (see e.g., Smith \& Arnett 2014), or perhaps by
some violent binary interaction like a collision or merger (Smith
2011; Smith \& Arnett 2014; Podsiadlowski et al.\ 2010).  Soker and
collaborators have discussed an accretion model to power the
luminosity in events like $\eta$ Car's Great Eruption, but these
assume that an eruption occurs to provide the mass that is then
accreted by a companion, and so there is no explanation for what
triggers the mass loss from the primary in the first place.  In any
case, research on these eruptions is actively ongoing; it is a major
unsolved problem in astrophysics, and in the study of VMSs in
particular.

\section{Very Luminous Supernovae}% - 10 pages
\label{sec:3}

\subsection{Background}

The recognition of a new regime of SN explosions has just occurred in
the last few years --- this includes SNe that are observed to be
substantially more luminous than a standard bright Type Ia SN (the
brightest among ``normal'' SNe). Although this is still a young field,
the implications for and connections to the evolution and fate of VMSs
is exciting.  Here we discuss these luminous SNe as well as gamma ray
bursts (GRBs), and their connection to the lives and deaths of the
most massive stars.

This field of research on the most luminous SNe took on a new
dimension with the discovery of SN~2006gy (Smith et al.\ 2007; Ofek et
al.\ 2007), which was the first of the so-called ``super-luminous
SNe'' (SLSNe).  The surprising thing about this object was that with
its high peak luminosity ($-$21.5 mag) and long duration (70 days to
rise to peak followed by a slow decline), the integrated luminous
energy $E_{rad}$ was a few times 10$^{51}$ erg, more than any previous
SN.  A number of other SLSNe have been discovered since then (see
below).  Why were these SLSNe not recognized previously? There may be
multiple reasons, but clearly one reason is that earlier systematic SN
searches were geared mainly toward maximizing the number of Type Ia SN
discoveries in order to use them for cosmology.  This meant that these
searches, which usually imaged one galaxy per pointing due to the
relatively small field of view, mainly targeted large galaxies to
maximize the chances of discovering SNe Ia each night.  Since it
appears that SLSNe actually seem to prefer dwarf galaxy hosts (either
because of lower metallicity, or because dwarf galaxies have higher
specific star-formation rates), these searches may have been biased
against discovering SLSNe.  More recent SN searches have used larger
fields of view and therefore search large areas of the sky, rather
than targeting individual large galaxies; this is probably the
dominant factor that led to the increased discovery rate of SLSNe (see
Quimby et al.\ 2011).  Additionally, even if SLSNe were discovered in
these earlier targeted searches, prescious followup resources for
spectroscopy on large telescopes are limited, and so SNe that were not
Type Ia were given lower priority.

%%%%%%%%%%%%%%%

\begin{figure}[!ht]
\includegraphics[scale=0.56]{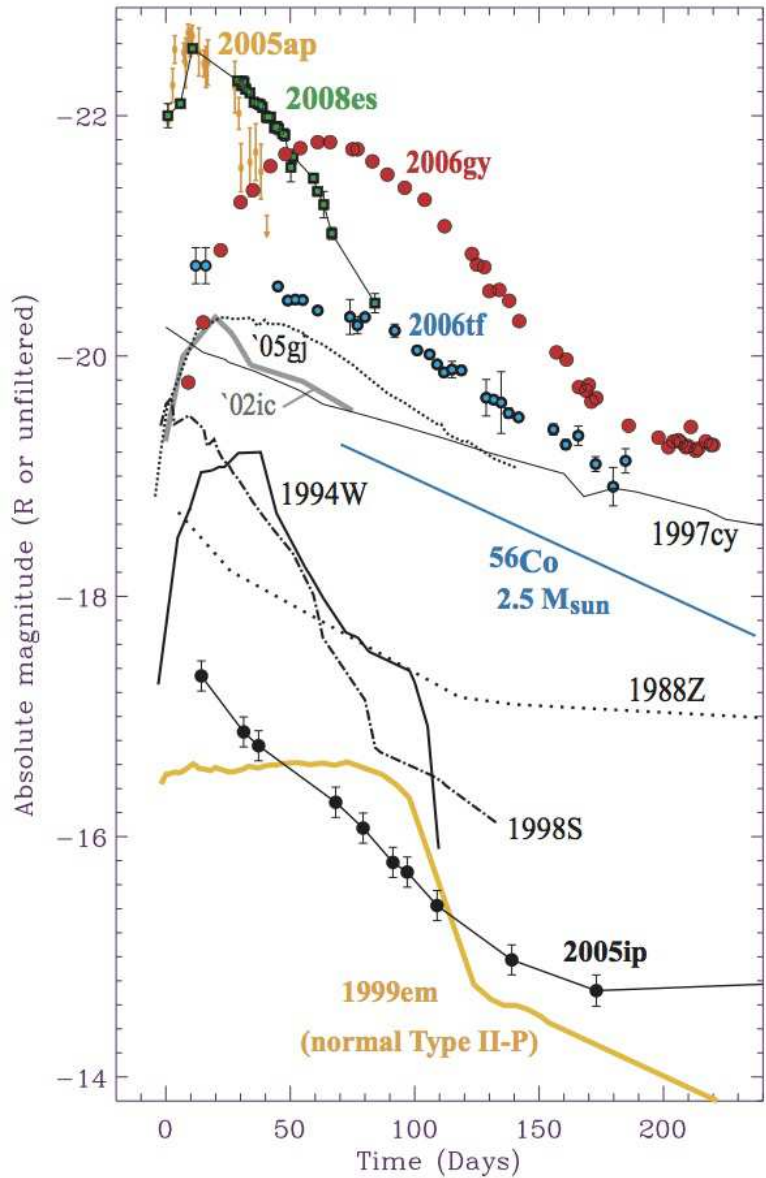}
\caption{Example light curves of several Type IIn SNe, along with two
  non-IIn SLSNe (SN~2005ap, a Type Ic) and SN~2008es (Type II) for
  comparison.  SN~1999em is also shown to illustrate a ``normal'' Type
  II-P light curve.  The fading rate of radioative decay from
  $^{56}$Co to $^{56}$Fe is indicated, although for most SNe~IIn this
  is not thought to be the power source despite a similar decline rate
  at late times in some objects. Note that SN~2002ic and SN~2005gl are
  thought to be examples of SNe Ia interacting with dense CSM, leading
  them to appear as Type IIn (see text).}
\label{fig:lc2n} 
\end{figure}

\subsection{Sources of unusually high luminosity}

So what can make SLSNe 10--100 times more luminous than normal SNe?
There are essentially two ways to get a very luminous explosion.  One
is by having a relatively large mass of $^{56}$Ni that can power the
SN with radioactive decay; a higher luminosity generally requires a
larger mass of synthesized $^{56}$Ni.  While a typical bright Type Ia
SN might have 0.5-1 $M_{\odot}$ of $^{56}$Ni, a super-luminous SN must
have 1-10 $M_{\odot}$ of $^{56}$Ni to power the observed luminosities.
Currently, the only proposed explosion mechanism that can do this is a
pair instability SN (see Chapter 7 by Woosley \& Heger).  It is
interesting to note that most normal SNe are powered by radioactive
decay -- were it not for the synthesis of $^{56}$Ni in these
explosions, we wouldn't ever see most SNe.

The synthesized mass of $^{56}$Ni needed to supply the luminosity of a
PISN through radioactivity is estimated from observations the same way
as for normal SNe:

\begin{equation}
L = 1.42 \times 10^{43} {\rm ergs \ s^{-1}} e^{-t/111d} \, M_{\rm Ni}/M_{\odot}
\end{equation}

\noindent (Sutherland \& Wheeler 1984) where $L$ is the bolometric
luminosity at time $t$ after explosion (usually measured at later
times when the SN is clearly on the radioactive decay tail).
Important uncertainties here are that $L$ must be the {\it bolometric}
luminosity, which is not always easily obtained without good
multiwavelength data (otherwise this provides only a lower limit to
the $^{56}$Ni mass), and the time of explosion $t$ must be known (this
is often poorly constrained observationally, since most SNe have been
discovered near maximum luminosity).  An additional cause of ambiguity
is that in very luminous SNe, it is often difficult to determine if
the source of luminosity is indeed radioactivity, since other
mechanisms (see below) may be at work.

The other way to generate an extraordinarily high luminosity is to
convert kinetic energy into heat, and to radiate away this energy
before the ejecta can expand and cool adiabatically.  This mechanism
fails for many normal SNe, since the explosion of any progenitor star
with a compact radius (a white dwarf, compact He star, blue
supergiant) must expand to many times its initial radius before the
photosphere is large enough to provide a luminous display.  These SNe
are powered primarily by radioactivity, as noted above.  Red
supergiants, on the other hand, have larger initial radii, and so
their peak luminosity is powered to a much greater extent by radiation
from shock-deposited thermal energy.  However, even the bloated radii
of red supergiants (a few AU) are far smaller than a SN photosphere at
peak ($\sim$10$^{15}$ AU), and so the most common Type II-P SNe from
standard red supergiants never achieve an extraordinarily high
luminosity.  Most of the thermal energy initially deposited in the
envelope is converted to kinetic energy through adiabatic expansion.
This inefficiency (and relatively low $^{56}$Ni yields of only
$\sim$0.1 $M_{\odot}$) is why the total radiated energy of a normal
SN~II-P (typically 10$^{49}$ erg) is only about 1\% of the kinetic
energy in the SN ejecta.\footnote{Of course, most of the energy from a
  core collapse SN escapes in the form of neutrinos ($\sim$10$^{53}$
  erg).}

Smith \& McCray (2007) pointed out that this shock-deposition
mechanism could achieve the extremely high luminosities of SLSNe like
SN~2006gy if the initial ``stellar radius'' was of order 100 AU, where
this radius is not really the hydrostatic photospheric radius of the
star, but is instead the radius of an opaque CSM shell ejected by the
star before the SN.  The key in CSM interaction is that something else
(namely, pre-SN mass loss) has already done the work against gravity
to put a large mass of dense and slow-moving material out at large
radii ($\sim$10$^{15}$ cm) away from the star.  When the SN blast wave
crashes into this material, already located at a large radius, the
fast SN ejecta are decelerated and so the material is heated far from
the star, where it can radiate away its thermal energy before it
expands by a substantial factor.  By this mechanism, large fractions
($\sim$50\% or more) of the total ejecta kinetic energy can be
converted to thermal energy that is radiated away.  In a hydrogen-rich
medium, the photosphere tends to an apparent temperature around
6000-7000 K, and so a large fraction of the radiated luminosity
escapes as visual-wavelength photons.  Since this mechanism of
optically thick CSM interaction is very efficient at converting ejecta
kinetic energy into radiation, this process can yield a SLSN without
an extraordinarily high explosion energy or an exotic explosion
mechanism.  What makes this scenario extraordinary (and a challenge to
understand) is the requirement of ejecting $\sim$10 $M_{\odot}$ in
just the few years before core collapse.  This is discussed more
below.

A variant of this conversion of kinetic energy into light is powering
a SLSN with the birth of a magentar (Woosley 2010; Kasen \& Bildsten
2010).  In this scenario, a normal core-collapse SN explodes the star
and sends its envelope (10s of $M_{\odot}$) expanding away from the
star.  For the SN itself, there is initially nothing unusual compared
to normal SNe.  But in this case a magnetar is born instead of a
normal neutron star or black hole.  The rapid spin-down of the
magnetar subsequently injects $\sim$10$^{51}$ ergs of energy into the
SN ejecta (which have now expanded to a large radius of $\sim$100 AU).
Similar to the opaque CSM interaction model mentioned previously, this
mechanism reheats the ejected material at a large radius, so that it
can radiate away the energy before the heat is lost to adiabatic
expansion, providing an observer with a SLSN.  It would be very
difficult to tell the difference observationally between the magnetar
model and the opaque shocked shell model during the early phases
around peak when photons are diffusing out through the shell or
ejecta.  It may be possible to see the difference at late times if
late-time data are able to see the signature of the magnetar (Inserra
et al. 2013).

In summary, there are three proposed physical mechanisms for powering
SLSNe.  For each, there are also reasons to suspect a link to VMSs.

{\bf 1. Pair instability SNe.}  This is a very powerful thermonuclear
SN explosion.  To produce the observed luminosity and radiated energy,
one requires of order 10 $M_{\odot}$ of synthesized $^{56}$Ni.  These
explosions are only expected to occur in VMSs with initial masses of
$>$150 $M_{\odot}$, because those stars are the only ones with a
massive enough CO core to achieve the high temperatures needed for the
pair-instability mechanism.  The physics of these explosions is
discussed more in the chapter by Woosley \& Heger.  So far, there is
only one observed example of a SN that has been suggested as a good
example of a PISN, and this is SN~2007bi (Gal-Yam et al.\ 2009).
However, this association with a PISN is controversial. Dessart et
al.\ (2012) have argued that SN~2007bi does not match predictions for
a PISN; it has a very blue color with a peak in the UV, whereas the
very large mass of Fe-group elements in a PISN should cause severe
line blanketing, leading to very red observed colors and deep
absorption features.  Thus, it is unclear if we have ever yet observed
a PISN.

{\bf 2.  Opaque shocked shells.}  Here we have a normal SN explosion that
collides with a massive CSM shell, providing a very efficient way of
converting the SN ejecta kinetic energy into radiated luminosity when
the SN ejecta are decelerated.  The reason that this mechanims would
be linked to VMS progenitors is because one requires a very large mass
of CSM (10-20 $M_{\odot}$) in order to stop the SN ejecta.  Given
expectations for the minimum mass of SN ejecta in models and the fact
that stars also suffer strong mass loss duing their lifetimes, a high
mass progenitor star is needed for the mass budget.  Also, sudden
eruptive mass loss in non-terminal events that eject $\sim$10
$M_{\odot}$ is, so far, a phenomenon exclusivy associated with VMSs
like LBVs.  Although lower-mass stars do appear to be suffering
eruptions that look similar (see above), these do not involve the
ejection of 10 $M_{\odot}$.

{\bf 3.  Magnetar-powered SNe.}  In principle, the mechanism is quite
similar to the opaque shocked shell model, in the sense that thermal
energy is injected at a large radius, although here we have magnetar
energy being dumped into a SN envelope, rather than SN ejecta
colliding with CSM.  Although the SN explosion that leads to this SLSN
may be normal, the potential association with VMSs comes from the
magnetar.  Some magnetars have been found in the Milky Way to be
residing in massive young star clusters that appear to have a turnoff
mass around 40 $M_{\odot}$, suggesting that the progenitor of the
magnetar had an initial mass above 40 $M_{\odot}$.

\subsection{Type IIn SLSNe}

Since massive stars are subject to strong mass loss, it is common that
there is CSM surrounding a massive star at the time of its death, into
which the fast SN ejecta must expand.  The collision between the SN
blast wave and this CSM is referred to as ``CSM interaction'', which
is commonly observed in core-collapse SNe in the form of X-ray or
radio emission (Chevalier \& Fransson 1994).  However, only about
8-9\% of core-collapse SNe (Smith et al.\ 2011b) have CSM that is
dense enough to produce strong visual-wavelength emission lines and an
optically thick continuum.  In these cases, the SN usually exhibits a
smooth blue continuum with strong narrow H emission lines, and is
classified as a Type IIn SN.

Intense CSM interaction can occur in two basic regimes: (1) If the
interaction is optically thick so that photons must diffuse out
through the material in a time that is comparable to the expansion
time, or (2) an effectively optically thin regime, where luminosity
generated by CSM interaction escapes quickly.  This is equivalent to
cases where the outer boundary of the CSM is smaller or larger,
respectively, than the ``diffusion radius'' (see Chevalier \& Irwin
2011).  The former case will yield an observed SN without narrow
lines, resembling a normal broad-lined SN spectrum. The latter will
exhibit strong narrow emission lines with widths comparable to the
speed of the pre-shock CSM, emitted as the shock continues to plow
through the extended CSM.  In most cases, the SN will transition from
the optically thick case to the optically thin case around the time of
peak luminosity (see Smith et al.\ 2008).  If the CSM is hydrogen
rich, the narrow H lines earn the SN the designation of Type~IIn.  (If
the CSM is H-poor and He-rich, it will be seen as a Type Ibn, but
these are rare and no SLSNe have yet been seen of this type.)

Although narrow H emission lines are the defining charcteristic of the
Type~IIn class, the line widths and line profiles can be complex with
multiple components.  They exhibit wide diversity, and they evolve
with time during a given SN event as the optical depth drops and as
the shock encounters density and speed variations in the CSM.  These
line profiles are therefore a powerful probe of the pre-SN mass loss
from the SN prognitor star. Generally, the emission line profiles in
SNe~IIn break down into three subcomponents: narrow,
intermediate-width, and broad.

%------FIGURE 2006gy wind
\begin{figure}[b]
\includegraphics[scale=0.6]{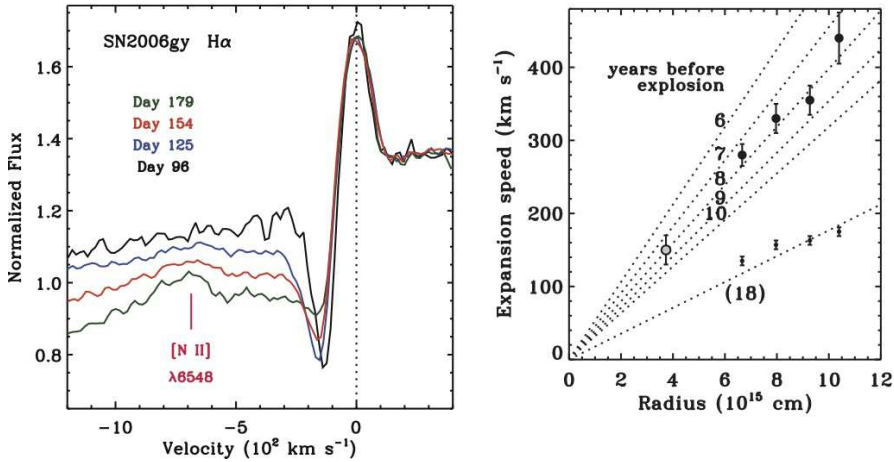}
\caption{Observations of pre-shock CSM speeds in the SLSN IIn
  SN~2006gy.  The left panel shows several tracings of the narrow P
  Cygni feature.  The right shows velocities measured for various
  radii, where dates have been converted to radii based on the
  observed expansion speed of the cold dense shell (upper points are
  for the blue edge of the absorption, while the lower points are for
  the velocity at the minimum of the absorption).  The CSM velocity
  follows a Hubble-like law, indicating a single ejection date for the
  CSM about 8 yr prior to the SN.  Both figures are from Smith et al.\
  (2010a).}
\label{fig:06gy} 
\end{figure}

\begin{itemize}

\item The narrow (few 10$^2$ km s$^{-1}$) emission lines arise from a
  photoionized shock precursor, when hard ionizing photons generated
  in the hot post-shock region propagate upstream and photoionize much
  slower pre-shock gas.  The width of the narrow component, if it is
  resolved in spectra, gives an estimate of the wind speed of the
  progenitor star in the years leading up to core collapse.  Since
  these speeds are generally between about 200--600 km s$^{-1}$, this
  seems to suggest blue supergiant stars or LBVs for the progenitors
  of SNe~IIn, because the escape speeds are about right (see SMith et
  al.\ 2007, 2008, 2010a).  Bloated red supergiants or compact WR
  stars have much slower or faster wind speeds, respectively.  In some
  cases when relatively high spectral resolution is used, one can
  observe the narrow P Cygni absorption profile.  This gives an even
  more precise probe of the wind speed of the pre-shock gas along the
  line-of-sight, which in some cases has multiple velocity components
  showing that the wind speed has been changing (see below, and Groh
  \& Vink 2011).  Since the absorption occurs in the densest gas
  immediately ahead of the shock, one can potentially use the time
  variation in the P Cyg absorption to trace out the radial velocity
  law in the wind.  A dramatic example of this was the case of
  SN~2006gy (Fig.~\ref{fig:06gy}; Smith et al.\ 2010a), where the
  velocity of the P~Cyg absorption increased with time as the shock
  expanded, indicating a Hubble-like flow in the CSM (i.e. $v \propto
  R$).  In this case, the Hubble-like law in the pre-shock CSM
  indicated that the dense CSM was ejected only about 8 yr before the
  SN (Smith et al.\ 2010a).  The rather close synchronization between
  the pre-SN eruptions and the SN has important implications, and is
  discussed more below.

\item Intermediate-width ($\sim$10$^3$ km s$^{-1}$) components usually
  accompany the narrow emission-line cores.  Generally these broader
  components exhibit a Lorentzian profile at early times and gradually
  transition to Gaussian, asymmetric, or irregular profiles at late
  times.  This is thought to be a direct consequence of dropping
  optical depth (see Smith et al.\ 2008).  At early times in very
  dense CSM, line photons emitted in the ionized pre-shock CSM must
  diffuse outward through optically thick material outside that
  region.  The multiple electron scatterings encountered as the
  photons escape produces the Lorentzian-shaped wings to the narrow
  line cores.  For these phases, it would therefore be a mistake to
  fit multiple components to the H$\alpha$ line profile, for example,
  and to adopt the broader component as indicative of some
  characteristic expansion speed in the explosion.  At later times
  when the pre-shock density is lower and we see deeper into the
  shock, the intermediate-width components can trace the kinematics of
  the post-shock region more directly.  These genrally indicate shock
  speeds of a few 10$^3$ km s$^{-1}$ or less.
 
\item Sometimes, in special cases of lower-density CSM (or at late
  times), clumpy CSM, or CSM with non-spherical geometry, one can also
  observe the broad-line profiles from the underlying fast SN ejecta.
  In these cases one can estimate the speed of the SN ejecta directly.
  This usually does not occur in the most luminous SNe~IIn, however,
  simply because the lower-density CSM or the small solid angle for
  CSM interaction (i.e. a disk) needed to allow one to see the broad
  SN ejecta lines also limits the luminosity of the CSM interaction,
  making it hard to have both transparency and high luminosity in the
  same explosion.  A recent case of this is teh 2012 SN event of
  SN~2009ip (Smith et al.\ 2014).

\end{itemize}

\subsection{CSM Mass Estimates for SLSNe IIn}

{\it Cold Dense Shell (CDS) Luminosity:} Armed with empirical
estimates of the speed of the CSM and the speed of the advancing
shock, one can then calculate a rough estimate for the density and
mass-loss rate of the CSM required to power the observed luminosity of
the SN.  Dense CSM slows the shock, and the resulting high densities
in the post-shock region allow the shock to become radiative.  With
high densities and optical depths, thermal energy is radiated away
primarily as visual-wavelength continuum emission.  This loss of
energy removes pressure support behind the forward shock, leading to a
very thin, dense, and rapidly cooling shell at the contact
discontinuity (usually referred to as the ``cold dense shell'', or
CDS; see Chugai et al.\ 2004; Chugai \& Danziger 1994).  This CDS is
pushed by ejecta entering the reverse shock, and it expands into the
CSM at a speed $V_{CDS}$.  In this scenario, the maximum emergent
continuum luminosity from CSM interaction is given by

\begin{equation}
L_{CSM} \ = \ \frac{1}{2} \, \dot{M} \, \frac{V_{CDS}^3}{V_W} \ = \ \frac{1}{2} \, w \, V_{CDS}^3
\end{equation}

\noindent where $V_{CDS}$ is the outward expansion speed of the CDS
derived from observations of the intermediate-width component, $V_W$
is the speed of the pre-shock wind derived from the narrow emission
line widths or the speed of the P Cygni absorption trough, $\dot{M}$
is the mass-loss rate of the progenitor's wind, and $w$ =
$\dot{M}/V_W$ is the so-called wind density parameter (see Chugai et
al.\ 2004; Chugai \& Danziger 1994; Smith et al.\ 2008, 2010a).  The
wind density parameter is a convenient way to describe the CSM
density, because it does not assume a constant speed (for the highest
mass-loss rates, it may be a poor assumption to adopt a constant wind
with a standard $R^{-2}$ density law, since the huge masses involved
are more likely to be the result of eruptive/explosive mass loss).

In general, this suggests that more luminous SNe require either higher
density in the CSM, faster shocks, or both.  Thus, a wide range of
different CSM density (resulting from different pre-SN eruption
parameters or different wind mass-loss rates) should produce a wide
variety of luminosities in SNe IIn.  This is, in fact observed.
Figure~\ref{fig:lc2n} shows several examples of light curves for
well-studied SNe IIn, which occupy a huge range in luminosity from the
most luminous SNe down to the lower bound of core-collapse SNe (below
peaks of about -15.5 mag, we would generally refer to a SN~IIn as a SN
impostor).

To derive a CSM mass, it is common to re-write the previous equation
with an efficiency factor $\epsilon$ as:

\begin{equation}
  L_{CSM} \ = \epsilon \ \frac{1}{2} \, \dot{M} \, \frac{V_{CDS}^3}{V_W} \ = \epsilon \  \frac{1}{2} \, w \, V_{CDS}^3.
\end{equation}

\noindent With representative values, this can be rewritten as:

\begin{equation}
  \dot{M} = 0.3 \, M_{\odot} \, {\rm yr}^{-1} \times \frac{L_9}{\epsilon^{-1}} \frac{V_w/200}{(V_{CDS}/2000)^3} 
\end{equation}

\noindent where $L_9$ is the bolometric luminosity in units of 10$^9$
$L_{\odot}$, $V_W/200$ is the CSM expansion speed relative to 200 km
s$^{-1}$, and $V_{CDS}/2000$ is the expansion speed of the post-shock
gas in the CDS relative to 2000 km s$^{-1}$.  These velocities are
representative of those observed in SNe~IIn, although there is
variation from one object to the next.  $L_9$ corresponds roughly to
an absolute magnitude of only $-$17.8 mag, which is relatively modest
for SNe~IIn (Fig.\ 3).  Thus, we see that even for relatively normal
luminosity SNe~IIn, extremely high pre-SN mass-loss rates are
required, much higher than is possible for any normal wind.  For SLSNe
that are $\sim$10 times more luminous, extreme mass-loss rates of
order $\sim$1 $M_{\odot}$ yr$^{-1}$ are needed.  Moreover, this
mass-loss rate is really a lower limit, due to the efficiency factor
$\epsilon$, which must be less than 100\%.  In favorable cases (fast
SN ejecta, slow and dense CSM) the efficiency can be quite high (above
50\%; see van Marle et al. 2010).  However, for lower densities and
especially non-spherical geometry in the CSM, the efficiency drops and
CSM mass requirements rise.

In cases where the post-shock H$\alpha$ emission is optically thin,
one can, in principle, also estimate the CSM mass in a similar way, by
replacing the bolometric luminosity with $L_{H\alpha}$, and the
efficiency $\epsilon$ with the corresponding H$\alpha$ efficiency
$\epsilon_{H\alpha}$.  This is perhaps most appropriate at late times,
as CSM interaction may contiunue for a decade after the SN.  During
this time the assumption of optically thin post-shock H$\alpha$
emission may be valid.  In practice, however, there are large
uncertainties in the value of $\epsilon_{H\alpha}$ (usually assumed to
be of order 0.005 to 0.05; e.g.\ Salamanca et al.\ 2002), so this
diagnostic provides only very rough order of magnitude estimates.

{\it Light Curve Fits:} The rough estimate in the previous method
provides a mass-loss rate corresponding only to the density overtaken
at one moment by the shock (assuming the CDS radiation escapes without
delay; see below).  In reality, the values of $V_{CDS}$, $V_{w}$, and
the CSM density can change with time as the shock decelerates while it
expands into the CSM, as does the speed of the SN ejecta crashing into
the reverse shock.  Moreover, pre-SN mass loss is likely to be
episodic, so it is unclear for how long that value of $\dot{M}$ was
sustained.  To get the total CSM mass ejected by the progenitor within
some time frame before core collapse (and hence, an average value of
$\dot{M}$), one must integrate over time.  This means producing a
model to fit the observed light curve.

One can calculate a simple analytic model for the CSM mass needed to
yield the light curve by demanding that momentum is conserved in the
collision between the SN ejecta and the CSM, and that the change in
kinetic energy resulting from the deceleration of the fast SN ejecta
is lost to radiation.  Assuming an explosion energy, a density law for
the SN ejecta, and a speed and density law of the CSM, one can
calculate the resulting analytic light curve assuming that high
densities and H-rich composition lead to a small bolometric correction
(see Smith et al.\ 2008, 2010a; Smith 2013a, 2013b; Chatzopoulos et
al.\ 2013; Moriya et al.\ 2013).  One can also do the same from a
numerical simulation (e.g., Woosley et al. 2007; van Marle et
al. 2010). In general, very high CSM masses of order 10-20 $M_{\odot}$
are found for SLSNe like SN~2006gy and 2006tf, emitted in the decade
or so preceeding the explosion.  Considering the CSM mass within the
radius ivertaken by the shock, the undertainty in this mass estimate
is roughyl a factor of 2, but should also be considered a lower limit
to the total mass since more mass can reside at larger radii.  When
very high mass and dense CSM is involved, this methos is usually more
reliable than other methods (emission lines, X-rays, radio) that may
severely underestimate the mass due to high optical depths.

{\it Diffusion Time:} In extreme cases where the CSM is very dense,
the diffusion time $\tau_{diff} \simeq (n \sigma R^2)c$ may be long.
If $\tau_{diff}$ becomes comparable to the expansion timescale of the
shock moving through the CSM $\tau_{exp} \simeq R/V_s$, then the
shock-deposited thermal energy can leak out after the shock has broken
out of the CSM.  Since the radius of the CSM may be very large (of
order 10$^{15}$ cm), this may produce an extremely luminous SN display
(Smith \& McCray 2007).  This is essentially the same mechanism as the
normal plateau luminosity of a SN~II-P (Falk \& Arnett 1977), but the
radius here is the radius of the CSM, not the hydrostatic radius of
the star.  This can be simplified to

\begin{equation}
  M_{CSM}/M_{\odot} \simeq R_{15} (\tau_{diff} / 23 days) 
\end{equation}

\noindent where $R_{15}$ is the assumed radius of the opaque CSM in
units of 10$^{15}$ cm, and $\tau_{diff}$ can be estimated from
observations of the characteristic fading time of the SN light curve.
Applying this to SLSNe like SN~2006gy yields a CSM mass of order 10-20
$M_{\odot}$ (Smith \& McCray 2007; Chevalier \& Irwin 2011).  This is
comparable to the estimates through the previous method.  The
underlying physical mechanism is the same as normal CSM interaction
discussed above, but the optical depths are assumed to be too high for
the luminosity to escape quickly.  In fact, even lower-luminosity
SNe~IIn may have diffusion-powered light curves at early times as the
shock breaks through the inner and denser parts of the wind; their
lower luminosity compared to SLSNe reflects the smaller radius in the
CSM where this breakout occurs (see Ofek et al.\ 2013a).

{\it H$\alpha$ Emission from Unshocked CSM:} When high resolution
spectra reveal a narrow P Cygni component to the H$\alpha$ line
(widths of order 100-500 km s$^{-1}$), one can infer that this
emission arises from the pre-shock CSM.  (Note that if a narrow P Cyg
profile is not seen, but rather a simple emission profile, it is
uncertain if this narrow component arises from a distant circumstellar
nebula or a nearby H~{\sc ii} region.)  Following Smith et al.\
(2007), the mass of emitting ionized hydrogen in the CSM around a
SN~IIn can be inferred from the total narrow-component H$\alpha$
luminosity $L_{H\alpha}$ from

\begin{equation}
M_{H\alpha} \simeq \frac{m_H L_{H\alpha}}{h \nu {\alpha}^{eff}_{H\alpha} n_e}
\end{equation}

\noindent where $h\nu$ is the energy of an H$\alpha$ photon,
$\alpha^{eff}_{H\alpha}$ is the Case B recombination coefficient, and
$n_e$ is the average electron density.  This simplifies to

\begin{equation}
M_{H\alpha} \simeq 11.4 \, M_{\odot} (L_{H\alpha}/n_e) 
\end{equation}

\noindent with $L_{H\alpha}$ in units of $L{\odot}$ and $n_e$ in
cm$^{-3}$ (see Smith et al.\ 2007).  Note that this is only the mass
of ionized H at high densities, so it is only a lower limit to the CSM
mass if some of the CSM remains neutral.  However, as with mass-loss
rates of normal O-type stars, the H$\alpha$ emission depends on the
degree of clumping in the wind (see review by Smith 2014), which can
lower the total required mass.  For more luminous SNe~IIn with very
dense pre-shock CSM, the narrow H$\alpha$ component mar arize from a
relatively thin zone ahead of the shock, and it therefore provides a
useful probe of the immediate pre-shock CSM in cases where a narrow P
Cyg profile is observed.  For a SLSN IIn like SN~2006gy, this method
yields a CSM mass of order 10 $M_{\odot}$ or a mass-loss rate of order
1 $M_{\odot}$ yr$^{-1}$ (Smith et al. 2007).  For a more
moderate-luminosity SN~IIn like SN~2009ip, Ofek et al.\ (2013a)
applied this same method and found a mass-loss rate of order 10$^{-2}$
$M_{\odot}$ yr$^{-1}$.

{\it X-ray and radio emission:} For SLSNe~IIn the X-ray and radio
emission is of limited utility in diagnosing the pre-SN mass-loss
rate, since very high CSM densities cause the X-rays to be self
absorbed (the reprocessing of X-rays and their thermalization to lower
temperatures is what powers the high visual-wavelength continuum
luminosity of SNe~IIn) and the CSM is optically thick to radio
emission during the main portion of the visual light curve peak.

When X-rays are detected, the X-ray luminosity $L_X$ can be used to
infer a characteristic mass-loss rate (see Ofek et al.\ 2013a; Smith
et al.\ 2007; Pooley et al. 2002):

\begin{equation}
L_X \simeq 3.8 \times 10^{41} ergs s^{-1} (\dot{M}/0.01)^2 (V_w/500)^{}-2 R_{15} e^{-(\tau + \tau_bf)}
\end{equation}

\noindent where $\dot{M}$ is in units of 0.01 $M_{\odot}$ yr$^{-1}$,
the wind speed is relative to 500 km s$^{-1}$, $R_{15}$ is the shock
radius in units of 10$^{15}$ cm, $\tau$ is the Thomson optical depth
in the wind, and the exponential term is due to wind absorption (see
Ofek et al.\ 2013a for further detail).  Caution must be used when
inferring global properties, however.  If the CSM is significantly
asymmetric (as most nebulae around massive stars are), X-rays may
indeed escape from less dense regions of the CSM/shock interaction,
while much denser zones may yield high optical depths and a strong
visual-wavelength continuum.  Thus, one could infer both low and high
densities simultaneously, which might seem contradictory at first
glance.  This was indeed the case in SN~2006gy, where the CSM density
indicated be X-rays was not nearly enough to provide the observed
visual luminosity (Smith et al.\ 2007).

Radio synchrotron emission is quashed for progenitor mass-loss rates
much higher than about 10$^{-5}$ $M_{\odot}$ yr$^{-1}$ in the first
year or so after explosion, and as a result, radio emission is rarely
seen from SNe~IIn at early times.  (In order for the CSM interaction
luminosity to compete with the normal SN photosphere luminosity, the
mass-loss rate of a SN~IIn progenior must generally be higher than
10$^{-4}$ $M_{\odot}$ yr$^{-1}$.  Moreover, very massive stars almost
always have normal winds in this range anyway, due to their high
luminosity.)  Radio emission can be detected at later times when the
density drops, but this emission is then tracing the mass-loss rate
that occurred centuries before the SN, rather than the eruptions in
the last few years before explosion. For a discussion of how to use
radio emission as a diagnostic of the progenitor's mass-loss rate, we
refer the reader to Chevalier \& Fransson (1994).

\subsection{Connecting SNe IIn and LBVs}

There are several lines of evidence that suggest a possible connection
between LBVs and the progenitors of SNe~IIn.  While each one is not
necessarily conclusive on its own, taken together they clearly favor
LBVs as the most likely known type of observed stars that fit the
bill.  If the progenitors of SNe~IIn are not actually LBVs, they do a
very good impersonation.  Here is a list of the different lines of
evidence that have been suggested:

(1) Super-luminous SNe IIn, where the demands on the amount of CSM
mass are so extreme (10-20 $M_{\odot}$ in some cases) that unstable
massive stars are required for the mass budget, and the inferred radii
and expansion speeds of the CSM require that it be ejected in an
eruptive event within just a few years before core collapse (Smith et
al.\ 2007, 2008, 2010a; Smith \& McCray 2007; Woosley et al.\ 2007;
van Marle et al.\ 2010).  So far, the only observed precedent for
stars known to exhibit this type of extreme, eruptive mass loss is LBV
giant eruptions.  (In fact, one could argue that since LBV is an
observational designation, if any such pre-SN event were to be
observed, we would probably call it an LBV-like eruption.)

(2) Direct detections of progenitors of SNe~IIn that are consistent
with massive LBV-like stars (Gal-Yam \& Leonard 2009; Gal-Yam et al.
2007; Smith et al. 2010b, 2011a, 2012; Kochanek et al.\ 2011).  This
is discussed in the next section (Sect. 4).

(3) Direct detections of non-terminal LBV-like eruptions preceding a
SN explosion.  This is seen by some as a smoking gun for an LBV/SN
connection.  So far there are only two clear cases of this, and two
more with less complete observations, discussed later (Sect. 5).

(4) The narrow emission-line components from the CSM indicate H-rich
ejecta surrounding the star.  H-rich CSM is obviously not exclusive to
LBVs, but it argues against most WR stars as the progenitors.  If SNe
IIn (especially SLSNe IIn) indeed require very massive progenitors,
this is a pretty severe problem for standard models of massive-star
evolution.  In any case, among massive stars with very strong mass
loss, LBVs are the only ones with the combination of H-rich ejecta and
high densities comparable to those required.

(5) Wind speeds consistent with LBVs.  As noted above, the observed
line widths for narrow components in SNe~IIn suggest wind speeds of a
few 10$^2$ km s$^{-1}$.  This is consistent with the expected escape
velocities of blue supergiants and LBVs (Salamanca et al.  2002; Smith
2006; Smith et al.\ 2007, 2008, 2010a; Trundle et al.\ 2008).  While
it doesn't prove that the progenitors are in fact LBVs, it is an
argument against red supergiants or WR stars as the likely
progenitors.  Wind speeds alone are not conclusive, however, since
radiation from the SN itself may accelerate pre-shock CSM to these
speeds.

(6) Wind variability that seems consistent with LBVs. Modulation in
radio light curves indicates density variations that suggest a
connection to the well-established variability of LBV winds (Kotak \&
Vink 2006).  Also, multiple velocity components along the line of
sight seen in blueshifted P Cygni absorption components of some SNe
IIn resemble similar multi-component absorption features seen in
classic LBVs like AG Car (Trundle et al.\ 2008).  This may hint that
some SN IIn progenitors had winds that transitioned across the
bistability jump, as do LBVs (see Vink chapter; Groh \& Vink 2011). As
with the previous point (wind speed), this is not a conclusive
connection to LBVs, since other stars do experience density and speed
variations in their winds, and the sudden impulse of radiation driving
from the SN luminosity itself might give the impression of multiple
wind speeds seen in absorption along the line of sight.  Nevertheless,
the variability inferred does hint at a possible connection to LBVs,
and is consistent with that interpretation.

%%%%

We must note, however, that not all SNe~IIn are necessarily tied to
LBVs and the most massive stars. Some SNe~IIn may actually be Type Ia
explosions with dense CSM (e.g., Silverman et al. 2013 and references
therein), some may be electron-capture SN explosions of stars with
initial masses around 8$-$10 $M_{\odot}$ (Smith 2013b; Mauerhan et
al.\ 2013b; Chugai et al.\ 2004), and some may arise from extreme red
supergiants like VY~CMa with very dense winds (Smith et al.\ 2009;
Mauerhan \& Smith 2012; Chugai \& Danziger 1994).  The argument for a
connection to LBVs and VMSs is most compelling for the SLSNe IIn
because of the required mass budget, which is hard to circumnavigate
(Smith \& McCray 2007; Smith et al.\ 2007, 2008, 2010a; Woosley et
al.\ 2007; Rest et al.\ 2011).

\subsection{Requirements for Pre-SN Eruptions and Implications}

In order for the characteristic Type~IIn spectrum to be observed, and
to achieve a high luminosity from CSM interaction, the collision
between the SN shock and the CSM must occur immediately after
explosion.  This places a strong constraint on the location of the CSM
and the time before the SN when it must have been ejected. Given the
luminosity of SLSNe, the photosphere must be at a radius of a few
10$^{15}$ cm, which must also be the location of the CSM if
interaction drives the observed luminosity.  Another way to arrive at
this same number is to require that a SN shock front (the cold dense
shell or CDS, as above) expands at a few 10$^3$ km s$^{-1}$ in order
to overtake the CSM in the first $\sim$100 days. Then we have $D$ = $v
\times t$ = (2,000 km s$^{-1}$) $\times$ (100 d) = 2$\times$10$^{15}$
cm.  Note that the observed blueshifted P Cygni absorption profiles in
narrow line components indicate that the CSM is {\it outflowing}.
This observed expansion rules out possible scenarios where the CSM is
primordial (i.e. disks left-over from star formation).

How recently was this CSM ejected by the progenitor star?  From the
widths of narrow lines observed in spectra we can derive the speed of
the pre-SN wind, and these show speeds of typically 100$-$600 km
s$^{-1}$ (Smith et al. 2008, 2010a; Kiewe et al. 2012), as noted
earlier.  In order to reach radii of 1-2$\times$10$^{15}$ cm, then,
the mass ejection must have occurred only a few years before the
SN. Since the lifetime of the star is several Myr and the time of He
burning is 0.5-1 Myr, a timescale of only 2-3 yr is very closely
synchronized with the time of core collapse. This is a strong hint
that something violent (i.e., hydrodynamic) may be happening to these
stars very shortly before core collapse, apparently as a {\it prelude}
to the core collapse event.

As noted earlier, the CSM mass must be substantial in order to provide
enough inertia to decelerate the fast SN ejecta and extract the
kinetic energy.  This is especially true for SLSNe, where high CSM
masses of order 10 $M_{\odot}$ are required.  Combined with the
expansion speeds of several 10$^2$ km s$^{-1}$ derived from narrow
emission lines in SNe~IIn, we find that whatever ejected the CSM must
have been provided with an energy of order 10$^{49}$ ergs.  Since the
mass loss occurred in only a few years before core collapse, it is
necessarily an eruptive event that is short in duration.

The H-rich composition, high mass, speed, and energy of these pre-SN
eruptions are remarkably similar to the physical conditions derived
for LBV giant eruptions.  This is the primary basis for the
connections between LBVs and SNe~IIn, as noted earlier.  Consequently,
we are left with the same ambiguity about the underlying physical
mechanism of pre-SN outbursts as we have for LBVs.  The SN precursors
seem to be some sort of eruptive or explosive mass-loss event, but the
underlying cause is not yet known.  Unlike many of the LBVs, however,
the pre-SN eruptions provide a telling clue --- i.e. for some reason
they appear to be synchronized with the time of core collapse.  This
is interesting, since we do know that core evolution proceeds rapidly
through several different durning stages as a massive star approaches
core collapse.  It is perhaps natural to associated these pre-SN
eruptions with Ne and O burning, each of which lasts roughly a year
(see Quataert \& Shiode 2011; Smith \& Arnett 2014).  Carbon burning
lasts at least several centuries (too long for the immediate SN
precursors, but possibly important in some SNe~IIn), while Si burning
lasts only a day or so (too short).  A number of possible
instabilities that may occur in massive stars during these phases is
discussed in more detail by Smith \& Arnett (2014), as well the
specific case of wave-driven mass loss by Quataert \& Shiode (2011)
and Shiode \& Quataert (2013).

In extremely massive stars with initial masses above $\sim$100
$M_{\odot}$, a series of precursor outbursts can occur as a result of
the pulsational pair instability (PPI; see chapter by Heger \&
Woosley).  These eruptions are thought to occur in a range of initial
masses (roughly 100-150 $M_{\odot}$) where explosive O burning events
are insufficient to completely disrupt the star as a final SN, but
which can give rise to mass ejections with roughly the mass and energy
required for conditions observed in luminous SNe IIn precursors.  The
PPI should occur far too rarely ($\sim$1\% or less of all
core-collapse SNe) to explain all of the SNe IIn (which are about
8-9\% of ccSNe; Smith et al. 2011b).  It may, however, provide a
plausible explanation for the much more rare cases of SLSNe of Type
IIn.

\subsection{Type Ic SLSNe and GRBs}

Not all SLSNe are Type~IIn, and not all SLSNe have H in their spectra.
The progenitors of SNe~IIn are required to eject a large mass of H in
just a few years before core collapse, so they must retain significant
amounts of H until the very ends of their lives.  This fact is in
direct conflict with stellar evolution models, as noted above.  There
are also, however, a number of SNe that may be associated with the
deaths of VMSs which have shed all of their H envelopes and possibly
their He envelopes as well before finally exploding.  Recall that SNe
with no visible sign of H, but which do show strong He lines are Type
Ib, and those which shown neither H or He are Type Ic (see Fillipenko
1997 for a review of SN classification).  (Type IIb is an intermediate
category that is basically a Type Ib, but with a small ($\sim$0.1
$M_{\odot}$) mass of residual H left, and so the SN is seen as a Type
II in the first few weeks, but then transitions to look like a Type
Ib.)  Together, Types Ib, Ic, and IIb are sometimes referred to as
``stripped envelope'' SNe.  The stripped envelope SNe most closely
related to the deaths of VMSs are the SLSN of Type Ic, and the
broad-lined Type Ic supernovae that are observed to be associated with
GRBs.

{\bf SLSN Ic.}  The most luminous SNe known to date turn out to be of
spectral Type Ic.  The prototypes for this class are objects like
SN~2005ap (Quimby et al.\ 2007) and a number of other cases discussed
by Quimby et al.\ (2011).  Although these SNe were discovered around
the same time as SN~2006gy, their true nature as the most luminous
Type Ic SNe wasn't recognized until a few years later.  This is
because they were actually located at a fairly substantial redshift
($z \simeq $0.2 to 0.3), causing their visual-wavelength spectra to
appear unfamiliar.  It turns out that these objects are closest to
Type Ic spectra, with no H and little if any He visible in their
spectra.\footnote{There is so far only one exception to this, which is
  SN~2008es (Miller et al.\ 2010; Gezari et al.\ 2010), whose light
  curve is shown in Figure~\ref{fig:lc2n}.  This object is a SLSN of
  Type II, with broad H lines in its spectra, and is not a Type IIn.
  The total mass of H in its envelope is not well constrained,
  however.}  Once their redshifts were recognized, it became clear
that these SNe were the most luminous of any SNe known, having peak
absolute magnitudes around $-$22 to $-$23.  These SNe are also hotter
than normal Type Ic SNe, however, with the peak of their spectral
energy distribution residing in the near-UV; this enhances their
visual-wavelength apparent brightness (and detectability) because of
the redshifts at which they are found.  The hotter photospheric
temperatures are likely related to their lack of H and He.  Unlike SNe
IIn, these obeject do not have narrow lines in their spectra; their
spectra exhibit broad absorption lines that are more like normal SNe.
More detailed information about these objects is available in two
recent reviews (Quimby et al.\ 2011; Gal-Yam 2012).

The three possible physical driving mechanisms for these explosions
are the same as those mentioned above for all SLSNe: (1) Interaction
between a SN shock and an opaque CSM shell, (2) Magnetar birth, or (3)
Pair instability SN.  Even though these objects do not have narrow
lines in their spectra (and therefore lack tell-tale signatures of CSM
interaction), the first is a possible power source if the opaque CSM
shell has a sharp outer boundary that is smaller than the diffusion
radius in the CSM.  If this is the case, then the shock will break out
of the CSM and photons will diffuse out afteward, producing a
broad-lined spectrum (Smith \& McCray 2007; Chevalier \& Irwin 2011).
Magnetar-driven SNe (Kasen \& Bildsten 2010; Woosley 2010) provide
another possible power source for SLSNe Ic, and so far appear to be
consistent with all available observations.  Recently, Inserra et al.\
(2013) have presented evidence that favors the magnetar model for
these SLSNe Ic, seen in the late-time data.  The third mechanism of a
pair instability SN (PISN) is perhaps the oldest viable idea for
making SLSNe from very massive stars (Barkatt et al.\ 1967; Bond et
al.\ 1984), but so far evidence for this type of explosion remains
unclear.  Most of the SLSNe Ic fade too quickly to be PISNe; for their
observed peak luminosities they would require $\sim$10 $M_{\odot}$ of
$^{56}$Ni in order to be powered by radioactive decay, but the rate at
which they are observed to fade from peak is much faster than the
$^{56}$Co rate (Quimby et al.\ 2011).  So far only one object among
the SLSNe Ic, SN~2007bi, has a fading rate that is consistent with
radioactivity (Gal-Yam et al.\ 2009), but the suggestion that this is
a true PISN is controversial, as noted earlier (Dessart et al.\ 2012).
It remains unclear if any PISN have yet been directly deteted.
Originally these SNe were predicted to occur only for extremely
massive stars in the early universe (with little mass loss), as
discussed more extensively in the chapter by Woosley \& Heger.

{\bf SNe Ic-BL associated with GRBs.} Gamma Ray Bursts (GRBs)
represent another example of the possible deaths of VMSs.  The
detailed observed properties of GRBS, the variety of GRBs (short
vs. long duration, etc.), and their history is too rich to discuss
here (see Woosley \& Bloom 2006 for a review).  Instead we focus on
the observable SNe that are associated with long-duration GRBs, which
are thought to result from core collapse to a black hole in the death
of a massive star.

So far, the only type of SN explosion seen to be associated with GRBs
are the so called ``broad-lined'' Type Ic, or SN~Ic-BL.  Here we must
be careful in terminology.  While earlier in this chapter we referred
to the fact that normal SNe have broad lines, at least compared to the
narrow and intermediate-width lines seen in SNe IIn, the class of
SN~Ic-BL have extremely broad absorption lines in their spectra.  A
normal SN typically has lines that indicate outflow speeds of
$\sim$10,000 km s$^{-1}$, but SNe Ic-BL exhibit expansion speeds
closer to 30,000 km s$^{-1}$, or 0.1c.  These trans-relativistic
speeds are related to the fact that a GRB has a highly relativistic
jet that is seen as the GRB, if we happen to be observing it nearly
pole-on.  Since kinetic energy goes as velocity squared, these very
fast expansion speeds in SNe Ic-BL imply large explosion energy, and
have led them to be referred to as ``hypernovae'' by some researchers.
The reason to associate these SNe Ic-BL and GRBs with the possible
deaths of VMSs is that the favored scenario for producing the
relativistic jet (the ``collapsar'', see Heger \& Woosley chapter)
involves a collapse to a black hole that is thought to occur in stars
with initial masses above 30 $M_{\odot}$.  Although the GRBs and the
afterglows are extremely luminous, the SN explosion seen as SNe Ic-BL
that follow the GRBs are not extremely luminous (they are near the top
end of the luminosity distribution for normal SNe, with peaks of $-$19
or $-$20 mag), and certainly not as luminous as the class of SLSN Ic
discussed above.

{\bf Host Galaxies.}  An interesting commonality is found between
SLSNe~Ic and the class of SN~Ic-BL associated with GRBs.  In addition
to sharing the Ic spectral type, indicating a progenitor stripped of
both its H and He layers, the two groups seem to arise preferentially
in similar environments. Namely, both classes of Ic occur
preferentially in relatively low-mass host galaxies with low
metallicity (Neill et al.\ 2011; Modjaz et al.\ 2008).  This may hint
that these two classes of SNe are the endpoints of similar evolution
in massive stars at low metallicity, but that some additional property
helps to determine if the object is a successful GRB or not.  Since
one normally associates stronger mass loss and stripping of the H and
He layers with stronger winds (and therefore higher metallicity), the
low-metallicity hosts of these SNe may hint that binary evolution
plays a key role in the angular momentum that is needed (especailly
for the production of GRB jets), with an alternative explanation
relying upon chemically homogeneous evolution of rapidly rotating
stars (see Yoon \& Langer 2005).  In this vein, it is perhaps
interesting to note that magnetars have been suggested as another
possible driving source for GRBs, while magnetar birth is also a
likely explanation for SLSNe Ic as noted above.  This is still an
active topic of current research.

\section{Detected Progenitors of Type IIn Supernovae}% - 5 pages
\label{sec:4}

While the previous section described inferred connections between very
luminous SNe and VMSs, these connections are however indirect, based
primarily on circumstantial evidence.  For example, they rely upon the
large mass of CSM needed in SNe IIn, the observed wind speeds, and the
requirement of extreme eruptive variability only demonstrated (to our
knowledge) by evolved massive stars like LBVs, and the possible
association with magnetars or collapsars.  However, our most direct
way to draw a connection between a SN and the mass of the star that
gave rise to it is to directly detect the progenitor star itself in
archival images of the explosion site taken before the SN occurred.
The increase of successful cases of this in recent years is thanks in
large part to the existence of archival {\it HST} images of nearby
galaxies, and this has now been done for a number of normal SNe and
for a small collection of Type~IIn explosions (only one of which
qualifies as a super-luminous SN).  For this technique to work in
identifying the progenitor star, one must be lucky\footnote{Another
  somewhat less direct technique for estimating the mass of a SN
  progenitor star is to analyze the stellar population in the nearby
  SN environment.  The age of the surrounding stellar population
  provides a likely (although not necessarily conclusive) estimate of
  the exploded star's lifetime and initial mass.  While this
  information can only be obtained for the nearest SNe, it can be
  performed after the SN fades and therefore does not require the
  lucky circumstance of having a pre-existing high-quality archival
  image.} enough to have a high quality, deep image of the explosion
site in a public archive.

The first cases of a direct detection of a SN progenitor star were the
very nearby explosions of SN~1987A in the Large Magellanic Cloud and
SN~1993J in M 81, using archival ground-based data.  With the advent
of {\it HST}, this technique could be pushed to host galaxies at
larger distances, and a number of such cases up until 2008 were
reviewed by Smartt (2009).  New examples continue to be added since
the Smartt (2009) review, including the very nearby SN IIb in M101,
SN~2011hd (Van Dyk et al.\ 2013).  Most of the progenitor detections
so far (and all those discussed in the Smartt 2009 review) are for SNe
II-P and IIb, all with relatively low implied initial masses ($<$20
$M_{\odot}$).

%%%%%

The technique for identifying SN progenitors requires very precise
work.  Once a nearby SN is discovered, one must determine if an
archival image of sufficient quality exists (it is frustrating, for
example, to find that your SN occurred at a position that is right at
the very edge of a CCD chip in an archival image, or just past that
edge).  Then one must obtain an {\it HST} image or high-quality
ground-based image (with either excellent seeing or adaptive optics)
of the SN itself, in order to perform very careful and precise
astrometry to pinpoint the exact position of the SN (usually the
precision is a few percent of an HST pixel).  The exact position of
the SN must then be identified on the pre-explosion archival image of
the SN site, using reference stars in common to both images
(preferably at the same wavelengths), and then finally one can
determine if there is a detected point source at the SN's location.
If not, one can derive an upper limit to the progenitor star's
luminosity and mass, which is most useful in the nearest cases where
the upper limit can be quite restrictive.  If there is a source
detected, then it becomes a ``candidate'' progenitor, because it could
also be a chance alignment, a companion star in a binary or triple
system, or a host cluster.  The way to tell is to wait several years
and verify that this candidate progenitor source has disappeared after
the SN fades beyond detectability.

Once a secure detection is made, one can then use the pre-explosion
image to estimate the apparent and absolute magnitudes of the star,
and to estimate colors if there are multiple filters.  After
correcting for the effects of extinction and reddening of the
progenitor (which might include the effects of unknown amounts of CSM
dust that was vaporized by the SN), one can place the progenitor star
on an HR diagram.  One can then use single-star evolution tracks to
infer a rough value for the star's initial mass, by comparing the
progenitor's position on the HR diagram to the expected luminosity and
temperatures at the endpoints of evolution models (note, however, that
trajectories of these evolution tracks are highly sensitive to
assumptions about mass loss and mixing in the models, and the 1D
models do not include possible instabilities in late burning phases;
see Smith \& Arnett 2014).  The technique favors types of progenitor
stars that are luminous in the filters used for other purposes
(usually nearby galaxy surveys, using $R$ and $I$-band filters),
allowing them to be more easily detected.  For example, WR stars are
the expected progenitors of at least some SNe~Ibc, but while these
stars are luminous, they are also hot and therefore emit most of their
flux in the UV.  Compared to a red supergiant at the same distance,
they are therefore less easily detected in the red $I$-band filters
that are often used in surveys of nearby galaxies that populate the
HST archive.  Similarly, very luminous progenitors that emit much of
their luminosity at visual eavelengths, like LBVs, should be
relatively easy to detect at a given distance.  This probably explains
why we have multiple cases of LBV-like progenitors, despite the
relatively small numbers of very massive stars.

%%%%%%%%

A central issue for understanding VMSs is whether they make normal SNe
when they die, rare and unusual types of SNe (like SLSNe or Type~IIn),
or if instead they have weak/failed SNe as core material and $^{56}$Ni
falls back into a black hole (making them difficult or impossible to
observe).  A common expectation from single-star evolution models
combined with core collapse studies (e.g., Heger et al.\ 2003 and
references therein; see also the chapter by Woosley \& Heger) is that
stars with initial masses above some threshold (for example, 30
$M_{\odot}$, although the exact value differs from one study to the
next) will collapse to a black hole and will fail to make a successful
bright SN explosion, unless special conditions such as very rapid
rotation and envelope stripping can lead to a collapsar and GRB.

Observationally, there are at least four cases where stars more
massive than 30 $M_{\odot}$ do seem to have exploded successfully, and
all of these are Type~IIn (recall that these cases may be biased
because LBV-like progenitos are very bright and easier to detect than
hotter stars of the same bolometric luminosity).  The four cases are
listed individually below.

{\bf SN~2005gl.} SN~2005gl was a moderately luminous SN~IIn (Gal-Yam
et al. 2007).  Pre-explosion images showed a source at the SN position
that faded below detection limits after the SN had faded (Gal-Yam \&
Leonard 2009). Its high luminosity suggested that the progenitor was a
massive LBV similar to P Cygni, with an initial mass of order 60
$M_{\odot}$ and a mass-loss rate shortly before core-collapse of
$\sim$0.01 $M_{\odot}$ yr$^{-1}$ (Gal-Yam et al.\ 2007).

{\bf SN~1961V.}  Another example of a claimed detection of a SN~IIn
progenitor, SN~1961V, has a more complicated history because it is
much closer to us and more highly scrutinized. For decades SN~1961V
was considered a prototype (although the most extreme case) of giant
eruptions of LBVs, as noted above, and an analog of the 19th century
eruption of $\eta$ Carinae (Goodrich et al.\ 1989; Filippenko et al.\
1995; Van Dyk et al.\ 2002). However, two recent studies (Smith et
al.\ 2011a; Kochanek et al.\ 2011) argue for different reasons that SN
1961V was probably a true core-collapse SN~IIn. Both studies point out
that the pre-1961 photometry of this source's variability was a
detection of a very luminous quiescent star, as well as a possible
precursor LBV-like giant eruption in the few years before the supposed
core collapse. While the explosion mechanism of SN~1961V is still
debated (e.g., Van Dyk \& Matheson 2012), the clear detection and
post-outburst fading of its LBV progenitor is at least as reliable as
the case for SN~2005gl.  SN~2005gl was shown to have faded to be about
1.5 mag fainter than its progenitor star, whereas SN~1961V is now at
least 6 mag fainter than its progenitor. In any case, the luminosity
of the progenitor of SN~1961V suggests an initial mass of at least
100-200 $M_{\odot}$.

In the previous two cases, the SN has now faded enough that it is
fainter than its detected progenitor star.  The implication is that
the luminous progenitor stars detected in pre-explosion images are no
longer there, and are likely dead.  This provides the strongest
available evidence that these detected sources were indeed the stars
that exploded to make the SNe we saw, and not simply a chance
alignment of another unrelated star, a star cluster, or a companion
star in a binary.  This is not true for the next two sources, which
are still in the process of fading from their explosion.  We will need
to wait until they fade to be sure that the candidate sources are
indeed the star that exploded.

{\bf SN~2010jl.} Of the four progenitor detections discussed here,
SN~2010jl is the only explosion that qualifies as a SLSN, with a peak
absolute magnitude brighter than $-$20 mag.  Smith et al.\ (2011c)
identified a source at the location of the SN in pre-explosion {\it
  HST} images.  The high luminosity and blue colors of the candidate
progenitor suggested either an extremely massive progenitor star or a
very young and massive star cluster; in either case it seems likely
that the progenitor had an initial mass well above 30 $M_{\odot}$.  In
this case, however, the SN has not yet faded (it is still bright after
3 yr), so we will need to wait to solve the issue of whether the
source was the prognitor or a likely host cluster.

{\bf SN~2009ip.}  Although its name says ``2009'', SN~2009ip is the
most recent addition to the class of direct SN~IIn progenitor
detections, because while the 2009 discovery event was a SN impostor,
the same object now appears to have suffered a true SN in 2012
(Mauerhan et al.\ 2013; Smith et al.\ 2014).  SN~2009ip is an
exceptional case, and is discussed in more detail below (Sect.\ 5).
For now, the relevant point to mention is that archival {\it HST}
images obtained a decade before the initial discovery revealed a
luminous point source at the precise location of the transient.  If
this was the quiescent progenitor star, the implied initial mass is
50-80 $M_{\odot}$ (Smith et al.\ 2010b) or $>$60 $M_{\odot}$ (Foley et
al.\ 2011), depending on the assumptions used to caculate the mass.
Thus, the case seems quite solid that the progenitor was indeed a VMS.

Altogether, all four of these cases of possible progenitors of SNe IIn
suggest progenitor stars that are much more massive than the typical
red supergiant progenitors of SNe II-P (Smartt 2009).

\section{Direct Detections of Pre-SN Eruptions} % - 8 pages
\label{sec:5}

SNe~IIn (and SNe Ibn) require eruptive or explosive mass loss in just
the few years preceding core collapse in order to have the dense CSM
needed for their narrow-line spectra and high luminosity from CSM
interaction.  As noted above, the timescale is constrained to be
within a few years beforehand, based on the observed expansion speed
of the pre-shock gas and the derived radius of the shock and
photosphere.

Until recently, these pre-SN eruptions were mostly hypothetical,
limited to conjectures supported by the circumstantial evidence that
{\it something} must deposit the outflowing CSM so close to the star.
However, we now have examples of SN explosions where a violent
outburst was detected in the few years before a SN, and in all cases
the SN had bright narrow emission lines indicative of CSM interaction.
The two most conclusive detections of an outburst are SN~2006jc and
SN~2009ip, and they deserve special mention.  SN~1961V and SN~2010mc
also had pre-peak detections, although the data are less complete, and
the interpretations are more controversial.

{\bf SN 2006jc.} - SN~2006jc was the first object clearly recognized
to have a brief outburst 2 years before a SN. The precursor event was
discovered in 2004 and noted as a possible LBV or SN impostor.  It had
a peak luminosity similar to that of $\eta$ Car (absolute magnitude of
$-$14), but was fairly brief and faded after only a few weeks
(Pastorello et al.\ 2007). No spectra were obtained for the precursor
transient source, but the SN explosion 2 years later was a Type Ibn
with strong narrow emission lines of He, indicating moderately slow
(1000 km s$^{-1}$) and dense H-poor CSM (Pastorello et al.\ 2007;
Foley et al.\ 2007).  There is no detection of the quiescent
progenitor, but the star is inferred to have been a WR star based on
the H-poor composition of the CSM.

%------FIGURE 2006gy wind
\begin{figure}[b]
\includegraphics[scale=0.58]{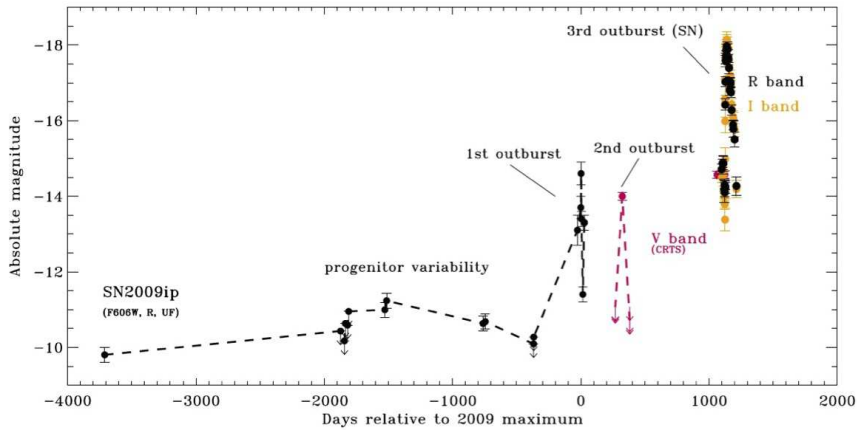}
\caption{The pre-SN light curve of SN~2009ip, from Mauerhan et al.\
  (2013).}
\label{fig:09ip} 
\end{figure}

{\bf SN 2009ip.} A much more vivid and well-documented case was
SN~2009ip, mentioned earlier. It was initially discovered and studied
in detail as an LBV-like outburst in 2009, again with a peak absolute
magnitude near $-$14.  This time, however, several spectra of the
pre-SN eruptions were obtained, and these spectra showed properties
similar to LBVs (Smith et al.\ 2010b; Foley et al.\ 2011). Also, a
quiescent progenitor star was detected in archival {\it HST} data
taken 10 yr earlier, which as noted above, indicated a VMS progenitor.
In the 5 yr preceding its discovery as an LBV-like eruption, the
progenitor also showed slow variability consistent with an S Dor-like
episode without a major increase in bolometric luminosity,
characteristic of LBVs. The object then experienced several brief
luminosity peaks over 3 yrs that looked like additional LBV eruptions
(unlike SN 2006jc, detailed spectra of these progenitor outbursts were
obtained), culminating in a final SN explosion in 2012 (Mauerhan et
al.\ 2013a; Smith et al.\ 2014). The SN light curve was double-peaked,
with an initially fainter bump ($-$15 mag) that had very broad (8000
km s$^{-1}$) emission lines probably formed in the SN ejecta
photosphere, and it rose quickly 40 days later to a peak of $-$18 mag,
when it looked like a normal SN~IIn (caused by CSM interaction, as the
SN crashed into the slow material ejected 1-3 years earlier; see
Mauerhan et al.\ 2013a and Smith et al.\ 2014). A number of detailed
studies of the bright 2012 transient have now been published, although
there has been some controversey about whether the 2012 event was a
true core-collapse SN (Mauerhan et al. 2013a; Prieto et al.\ 2013;
Ofek et al.\ 2013a; Smith et al.\ 2013, 2014) or not (Pastorello et
al.\ 2013; Fraser et al.\ 2013; Margutti et al.\ 2013).  More
recently, Smith et al.\ (2014) have shown that the object continues to
fade and its late-time emission is consistent with late-time CSM
interaction in normal Type IIn supernovae.  In any case, SN~2009ip
provides us with the most detailed information about any SN progenitor
for a decade preceding the SN, with a detection of a quiescent
progenitor, several LBV-like precursor eruptions of two different
types, and detailed high-quality spectra of the star.  This object
paints a very detailed picture of the violent death throes in the
final years in the life of a VMS.

{\bf SN~2010mc.}  Ofek et al.\ (2013b) reported the discovery of a
precursor outburst in the $\sim$40 days before the peak of SN~2010mc,
recognized after the SN by analyzing archival data.  Smith et al.\
(2013a,2013b) showed that the light curve of SN~2010mc was nearly
identical to that of the 2012 supernova-like event of SN~2009ip, to a
surprising degree.  Smith et al.\ (2014) proposed that the $\sim$40
day precursor events in both SN~2009ip and SN~2010mc were in fact the
SN explosions, since this is when very broad P Cygni features were
seen in the spectra, and that the following rise to peak was actually
due to additional luminosity generated by intense CSM interaction.  In
that case, the $\sim$40 day precursor event of SN~2010mc was not
actually a pre-SN eruption, but the SN itself.  Nevertheless, the
similarity in light curves and spectra between SN~2009ip and SN~2010mc
would obviously suggest that SN~2010mc probably did have a series of
pre-SN LBV-like erutions too, although those preceding events were not
detected.

{\bf SN~1961V.}  The remarkable object SN~1961V has extensive temporal
coverage of its pre-SN phases and solid detections of a luminous and
highly variable progenitor, moreso than any other SN.  The luminous
($-$12.2 mag absolute at blue/photographic wavelengths) progenitor is
well detected in data reaching back to more than 20 yr preceding the
SN, which includes some small ($\sim$0.5 mag) fluctuations in
brightness that could be S Dor-like LBV episodes.  In the year before
the SN, there is one detection at an absolute magnitude of roughly
$-$14.5, although since it is only one epoch, we don't know if this
was an LBV giant eruption or the beginning of the SN.  Then in 1961
there was a $\sim$100 day plateau at almost $-$17 mag followed by a
brief peak at about $-$18 mag.  After this, the SN faded rapidly and
has been fading ever since, except for some plateaus or humps in the
declining light curve within $\sim$5 yr after peak.  Currently, the
suggested source at the same position is about 6 mag fainter than the
progenitor, and shows H$\alpha$ emission.  In chronological order,
SN~1961V was therefore the first direct detection of a pre-SN
eruption.  In practice, however, the significance of this has been
overlooked because the 1961 event was discussed in terms of LBV
eruptions (it was considered a ``super-$\eta$ Car-like event''), and
was not thought to be a true SN.  It is only the much more recent
recognition that SN~1961V could have been a true core-collapse
Type~IIn supernova (Smith et al.\ 2011a; Kochanek et al.\ 2011) that
underscores the implications of the pre-1961 photometric evidence.

These direct discoveries of pre-SN transient events provide strong
evidence that VMSs suffer violent instabilities associated with the
latest phases in a massive star's life.  The extremely short timescale
of only a few years probably hints at severe instability in the final
nuclear burning sequences, especially Ne and O burning (Smith \&
Arnett 2014; Shiode \& Quataert 2013; Quataert \& Shiode 2012), each
of which lasts about 1 yr.  These instabilities may be exacerbated in
the most massive stars, athough much theoretical work remains to be
done.  The increased instability at very high initial masses is
certainly true in cases where the pre-SN eruptions result from the
pulsational pair instability (see the chapter by Woosley \& Heger),
but it may extend to other unkown nuclear burning instabilities as
well (Smith \& Arnett 2014).  Although the events listed above are
just a few very lucky cases, they may also be merely the tip of the
iceberg.  Undoubtedly, continued work on the flood of new transient
discoveries will reveal more of these cases.  Future cases will be
interesting if high-quality data can place reliable constraints on the
duration, number, or luminosity of the pre-SN outbursts that will
allow for a meaningful comparison with LBV-like eruoptions.  The
limitation will be the existence of high-quality archival data over
long timescales of years before the SNe, but these sorts of archives
are always becoming more populated and improved.  When LSST arrives,
it will probably become routine to detect pre-SN outbursts.

\section{Looking Forward (or Backward, Actually)} %- 2 pages
\label{sec:5}

Very massive stars are very bright, and their SLSNe are even brighter.
Thus, we can see them at large distances, and there is hope that we
may soon be able to see light from the explosions of some of the
earliest stars in the Universe.  The fact that VMSs appear to suffer
pre-SN instability that leads to the ejection of large amounts of mass
--- which in turn enhances the luminosity of the explosion --- helps
our chances of seeing the first SNe.  There is an expectation that the
low metallicity environments in the early Universe may favor the
formation of very massive stars because of the difficulty in cooling
and fragmentation during the star-formation process.

So then we must ask what happens to these stars and their explosions
as we move to very low metallicity?  How does the physics of eruptions
and explosions in the local universe translate to the low-metallicity
environments of the earlier universe?

Traditional expectations for massive star evolution are that lower
metallicity means lower mass-loss rates (e.g., Heger et al.\ 2003),
since line-driven winds of hot stars have a strong metallicity
dependence.  It is somewhat ironic, then, that the SNe associated with
VMSs have some of the most extreme mass-loss rates (SNe~IIn and SLSNe
Ic), {\it but these appear to favor host galaxies with low
  metallicity}.  This contradicts the simple expectation that lower
metallicity means lower mass loss, and the implication is that
eruptive mass-loss and mass transfer in binary systems may play an
extremely important role.  It may, in fact, dominate the observed
populations of different types of SNe (Smith et al.\ 2011b).  In that
case, extrapolating back to low-metallicity conditions in the early
univere is not so easy.  Binary evolution is not well understood even
in the local universe, so extrapolating to a regime where there is no
data remains rather adventurous.

The main theme throughout this chapter is that VMSs seem to suffer
violent eruptions that impact their evolution and drastically modify
the type of SN seen.  These eruptions may be very important and may
actually dominate the mass lost by VMSs in the local universe, and it
is important to recognize that they are probably much less sensitive
to changes in metallicity than line-driven winds.  The two leading
candidates for the physical mechanism of driving this eruptive mass
loss are continuum-driven super-Eddington winds and hydrodynamic
explosions.  While we are not yet certain of the triggering
mechanism(s) for either type of event, which may turn out to depend
somehow on metallicity, the {\it physical mechamisms} that drive the
mass loss are not metallicity dependent.

Super-Eddington continuum-driven winds rely on electron scattering
opacity to transfer radiation momentum to the gas (see the chapter by
Owocki; Owocki et al.\ 2004; Smith \& Owocki 2006), and this is
independent of metallicity since it only requires electrons supplied
by ionized H.  This occurs because absorption lines are saturated for
high densities in winds with mass-loss rates much above 10$^{-4}$
$M_{\odot}$ yr$^{-1}$ (recall that LBV eruptions typically have
mass-loss rates of 0.01 $M_{\odot}$ yr$^{-1}$ or more).  Non-terminal
hydrodynamic explosions are driven by a shock wave, and shock waves
can obviously still accelerate gas even with zero metal content.  If
the shocks are driven by some sort of instability in advanced nuclear
burning stages (using the ashes of previous burning stages as fuel),
it seems unlikely that this would depend sensitively on the initial
metallicity that the star was born with.  Since these eruptive
mechanisms appear to be important for heavy mass loss of VMS in the
local universe, there is a good chance that they will still operate or
may even be enhanced at low metallicity (Smith \& Owocki 2006).  The
recent recognition that SLSNe appear to favor low-metallicity hosts
(see above) would seem to reinforce this suspicion.

One of the key missions for the {\it James Webb Space Telescope} ({\it
  JWST}) will be to detect the light of the explosions from the first
stars.  Given the arguments above, we should perhaps be hopeful that
{\it JWST} may be able to see extremely luminous SNe from very massive
stars, if they suffer similar types of pre-SN eruptive mass loss.

\end{document}